\documentclass[creativecommons,sharealike,noncommercial,copyright]{eptcs}

\providecommand{\event}{MARS 2024}

\usepackage{url}
\usepackage{breakurl}

\usepackage{soul}  

\usepackage{listings}
\usepackage{mcrl}
\usepackage{mcrl2}
\usepackage{cadp-lotos}
\usepackage{cadp-lnt}

\usepackage{xcolor}

\usepackage{tikz}
\usetikzlibrary{patterns,arrows.meta}

\usepackage{amsmath}

\definecolor{grey}{cmyk}{0,0,0,0.8}
\definecolor{gray}{cmyk}{0,0,0,0.5}
\hypersetup{bookmarks=true,bookmarksopen=false,colorlinks=true,linkcolor=grey,citecolor=grey,urlcolor=grey}

\newcommand{\B}[1]{\mbox{\bf #1}}

\newcommand{\T}[1]{\mbox{\tt #1}}

\newcommand{\TRANS}{\textsc{Trans}}
\newcommand{\LINK}{\textsc{Link}}
\newcommand{\PHY}{\textsc{Phy}}
\newcommand{\CABLE}{\textsc{Cable}}
\newcommand{\BUS}{\textsc{Bus}}
\newcommand{\APPLI}{\textsc{Appli}}
\newcommand{\NODE}{\textsc{Node}}
\newcommand{\MAIN}{\textsc{Main}}

\newcommand{\ssig}{\texttt{Start}}
\newcommand{\esig}{\texttt{End}}
\newcommand{\subactgap}{\texttt{subactgap}}
\newcommand{\ackrec}{\texttt{ackrec}}
\newcommand{\ackmiss}{\texttt{ackmiss}}
\newcommand{\prefix}{\texttt{Prefix}}
\newcommand{\won}{\texttt{won}}
\newcommand{\lost}{\texttt{lost}}
\newcommand{\release}{\texttt{release}}
\newcommand{\hold}{\texttt{hold}}
\newcommand{\fair}{\texttt{fair}}
\newcommand{\immediat}{\texttt{immediate}} 

\newenvironment{blue}{\color{blue}}{\color{black}}
\newenvironment{red}{\color{red}}{\color{black}}

\newcommand{\IGNORE}[1]{}

\title{Four Formal Models of IEEE 1394 Link Layer}

\def\titlerunning{Four Formal Models of IEEE 1394 Link Layer}

\author{
   Hubert Garavel
   \institute{Univ. Grenoble Alpes, {\sc Inria}, {\sc Cnrs}, Grenoble {\sc Inp},
              {\sc Lig}, 38000 Grenoble, France}
   \email{hubert.garavel@inria.fr}
\and
   Bas Luttik
   \institute{Eindhoven University of Technology, The Netherlands}
   \email{s.p.luttik@tue.nl}
}

\def\authorrunning{H.~Garavel \& B.~Luttik}

\begin{document}

\maketitle

\begin{abstract}
We revisit the IEEE~1394 high-performance serial bus (``FireWire''), which
became a success story in formal methods after three PhD students, by using
process algebra and model checking, detected a deadlock error in this IEEE
standard. We present four formal models for the asynchronous mode of the
Link Layer of IEEE~1394: the original model in $\mu$CRL, a simplified model
in mCRL2, a revised model in LOTOS, and a novel model in LNT.
\end{abstract}



\section{Introduction}
\label{INTRO}

IEEE~1394 (also called ``FireWire'') is an interface standard that specifies
a serial bus architecture for high-speed communications. It can connect up
to 63~peripherals in a tree or daisy-chain topology, and can perform both
asynchronous and isochronous transfers simultaneously. It was developed
between 1986 and 1995 by a large consortium gathering Apple, Panasonic,
Philips, Sony, and many others contributors. This work resulted in an IEEE
standard~\cite{IEEE-1394}, followed by integration in many industrial
products.

In the framework of the COST-247 action \cite{COST-247}, a pan-European
academic collaboration that took place between 1994 and 1997, the asynchronous
mode of the link layer protocol of IEEE~1394 was selected as an interesting
case study for formal methods.
This protocol, which was close to being standardized, was thus
studied by several young scientists at this time. At CWI Amsterdam, Bas
Luttik developed a formal model~\cite{Luttik-97-a,Luttik-97-b} in the $\mu$CRL
language~\cite{Groote-Ponse-95,Groote-97} and stated five correctness
properties that the protocol should satisfy. At INRIA Grenoble, Mihaela
Sighireanu translated this model to LOTOS~\cite{ISO-8807} and, using the XTL
model checker \cite{Mateescu-Garavel-98} with the help of Radu Mateescu,
discovered that the deadlock-freeness property did not hold, i.e., that
the protocol could enter a deadlock state after a specific sequence of
50~transitions
\cite{Sighireanu-Mateescu-97-a,Sighireanu-Mateescu-97-b,Sighireanu-Mateescu-98}.
A detailed report about this bug, which would have been difficult to detect
using step-by-step simulation or testing, can be found in \cite{Firewire-98}.
The link layer protocol was also studied using theorem proving at the
Universities of Kiel and Eindhoven by Lars K\"uhne, Jozef Hooman, and
Willem-Paul de Roever \cite{Kuhne-Hooman-deRoever-97}.

Although the IEEE~1934 serial bus is no longer used today (deployed in Apple
products from 1999 to 2016, it has been gradually replaced by USB~2, USB~3,
and Gigabit Ethernet), it is an inspiring example for the formal methods
community. From a historical perspective, it is a striking success
story where three doctoral students discovered in a few weeks an unexpected
deadlock in an IEEE~standard designed and scrutinized over ten years by one
hundred experts. Also, numerous research papers have been devoted to another
aspect of IEEE~1934, its leader election algorithm (``root contention
protocol''), the verification of which involves parameters, probabilities,
and real time \cite{Shankland-vanderZwaag-98, Romijn-99-b,Shankland-Verdejo-99,
Stoelinga-Vaandrager-99,Devillers-Griffioen-Romijn-Vaandrager-00,
Maharaj-Shankland-00,Verdejo-Pita-MartiOliet-00,Romijn-01,Shankland-Verdejo-01,
Simons-Stoelinga-01,Daws-Kwiatkowska-Norman-02,Carchiolo-Malgeri-Mangioni-03-b,
Kwiatkowska-Norman-Sproston-03,Langevelde-Romijn-Goga-03,Romijn-03,Stoelinga-03,
Verdejo-Pita-MartiOliet-03,Daws-Kwiatkowska-Norman-04}.

Concerning the link layer protocol, formal methods evolved since 1997, as the
$\mu$CRL and LOTOS languages have been replaced by newer languages, respectively
mCRL2 \cite{Groote-Mathijssen-Weerdenburg-Usenko-06,
Groote-Mathijssen-Reniers-et-al-07,Groote-Mousavi-14,
Bunte-Groote-Keiren-et-al-19} and LNT \cite{Garavel-Sighireanu-98-a,Garavel-07,
Garavel-Lang-Serwe-17,Traian-3.12,Champelovier-Clerc-Garavel-et-al-10-v7.2},
a descendent of the E-LOTOS standard \cite{ISO-15437}.
Therefore, twenty-five years after, we revisit this case study to present,
along with the original $\mu$CRL model, three companion models:
a model written in mCRL2 by Jan Friso Groote, a recent revision of the LOTOS
model developed by M.~Sighireanu, and a novel model written in LNT.

The present article is organized as follows. Section~\ref{FIREWIRE} gives an
overview of the IEEE~1394 architecture and explains the behaviour of the Link
layer and neighbour layers. Section~\ref{MODELS} presents four formal models
in $\mu$CRL, mCRL2, LOTOS, and LNT, and discusses their main features from
a modelling point of view --- the models themselves being fully provided
in Annexes~\ref{ANNEX-MCRL} to \ref{ANNEX-LNT}. Section~\ref{VERIFICATION}
briefly reports about the verification (model checking and equivalence
checking) done on these models. Finally, Section~\ref{CONCLUSION} gives a few
concluding remarks.


\section{IEEE~1394 bus}
\label{FIREWIRE}

In this section, we present a description of IEEE~1394 that bridges the gap
between the general description given in the IEEE standard \cite{IEEE-1394}
and the four formal models provided in the present article.
The text in this section is based upon the technical report \cite{Luttik-97-a}
in which the $\mu$CRL model first appeared --- actually, this model was
developed from a draft version \cite{IEEE-1394-draft} of the IEEE standard,
but we believe that there is no significant difference between the draft and 
the standard in this respect.

First, we present the architecture as
defined in the standard. Then, we focus our attention on the link
layer of the protocol, the behaviour of which is our primary
modelling purpose. To provide a comprehensive description of the link
layer interacting with its environment, we will need to include the external 
functional behaviour of the physical layer, and so that is described too.


\subsection{Architecture}
\label{ARCHITECTURE}

The IEEE~1394 standard deals both with the physical requirements
and the protocol of the bus. The main feature of the standard is that it
supports two modes of transaction: an \emph{asynchronous mode}  and an
\emph{isochronous mode}.

In asynchronous mode, one party (the sender)
can send a message of arbitrary length to some other party (the
receiver). Such a message may be sent at an arbitrary moment after the
sender has gained access to the bus; the only timing restriction is
that the interval during which a node may have access to the bus is
bounded. In this mode, the receiver must confirm the receipt of the
message by sending an acknowledgement.

In isochronous mode the sender is obliged to send messages at fixed
rates, and messages are not acknowledged. This service is useful for
fast transmission of large amounts of data (e.g., audio/video streams),
if certainty at the side of the sender about the receipt of the data by
the receiver is not important, whereas the arrival of the data at a constant
rate is.

\begin{figure}[htb]
\begin{center}
\begin{tikzpicture}[rounded corners,inner sep=2pt,scale=1.5]
\draw[line width=5pt] (-3.5,0) -- (3.5,0);

\draw[fill=gray] (-3.5,1) rectangle  (-1.7,2.8);
\draw[fill=gray] (-1.5,1) rectangle  (0.3,2.8);
\draw[fill=gray] (1.5,1) rectangle  (3.3,2.8);

\draw[line width=5pt] (-2.5,0) -- (-2.5,1.3);
\draw[line width=5pt] (-0.5,0) -- (-0.5,1.3);
\draw[line width=5pt] (2.5,0) -- (2.5,1.3);

\draw (0,-0.2) node{\CABLE};

\draw (-2.5,1.3) node[sharp corners,rectangle,fill=white, minimum width=1.4cm] {\PHY{}};
\draw (-2.5,1.7) node[sharp corners,rectangle,fill=white,minimum width=1.4cm] {\LINK{}};
\draw (-2.5,2.1) node[sharp corners,rectangle,fill=white,minimum width=1.4cm] {\TRANS};
\draw (-3.2,1.7) node[sharp corners,rectangle,fill=white,minimum width=1.4cm,rotate=90] {\tiny CONTROLLER};

\draw (-2.6,2.6) node {\textcolor{white}{\small node $1$}};

\draw (-0.5,1.3) node[sharp corners,rectangle,fill=white, minimum width=1.4cm] {\PHY{}};
\draw (-0.5,1.7) node[sharp corners,rectangle,fill=white,minimum width=1.4cm] {\LINK{}};
\draw (-0.5,2.1) node[sharp corners,rectangle,fill=white,minimum width=1.4cm] {\TRANS};
\draw (-1.2,1.7) node[sharp corners,rectangle,fill=white,minimum width=1.4cm,rotate=90] {\tiny CONTROLLER};

\draw (-0.6,2.6) node {\textcolor{white}{\small node $2$}};

\draw (2.5,1.3) node[sharp corners,rectangle,fill=white, minimum width=1.4cm] {\PHY{}};
\draw (2.5,1.7) node[sharp corners,rectangle,fill=white,minimum width=1.4cm] {\LINK{}};
\draw (2.5,2.1) node[sharp corners,rectangle,fill=white,minimum width=1.4cm] {\TRANS};
\draw (1.8,1.7) node[sharp corners,rectangle,fill=white,minimum width=1.4cm,rotate=90] {\tiny CONTROLLER};

\draw (2.4,2.6) node {\textcolor{white}{\small node $n$}};

\draw[fill=black] (0.6,2) circle [radius=2pt];
\draw[fill=black] (0.9,2) circle [radius=2pt];
\draw[fill=black] (1.2,2) circle [radius=2pt];
\end{tikzpicture}
\end{center}
\caption{IEEE 1394 architecture}\label{fig:architecture}
\end{figure}
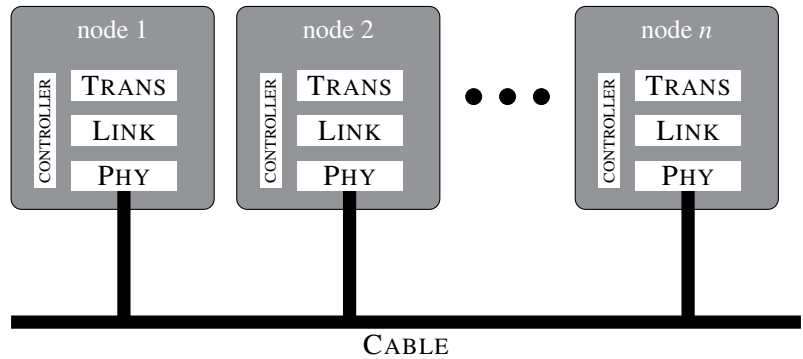

The IEEE~1394 serial bus architecture is roughly as depicted in
Figure~\ref{fig:architecture}.  It consists of a number of nodes
(addressable entities that run their own part of the protocol)
connected by a serial cable.

The protocol describing the behaviour of a node in asynchronous mode
distinguishes three layers:
\begin{enumerate}
  \item The \emph{transaction layer} (the upper layer, indicated by
    \TRANS{} in Figure~\ref{fig:architecture}) offers three types of
    transactions to the application(s) running on the node:
    \emph{read transactions} (read data from another node),
    \emph{write transactions} (write data to another node), and
    \emph{lock transactions} (have some of its own data
    processed by another node after which it is transferred
    back). Such transactions consist of a request and a response; the
    transaction layer can both handle \emph{concatenated response}
    transactions (response follows request immediately) and
    \emph{split} transactions (response not necessarily follows
    immediately on the request it belongs to).
  \item
    The \emph{link layer} (the middle layer, indicated by \LINK{} in
    Figure~\ref{fig:architecture}) forms the interface between the
    transaction layer and the physical components of the bus
    (consisting of the physical layers, which are connected to each
    other by a serial cable). The link layer provides two types
    of services to the transaction layer:

    \begin{description}
      \item[Data request/response:] By means of a \emph{\LINK{} data
        request}, the transaction layer instructs the link layer to send a packet to some particular
        node or to broadcast a packet to all other nodes. The
        transaction layer must react on a packet addressed to it by sending an
        acknowledge packet by means of a \emph{\LINK{} data response}.
        \item[Data indication/confirmation:] By  means of a \emph{\LINK{}
          data indication}, the link layer indicates the arrival of
        data (either request or response data).  The receipt of an
        acknowledge packet is indicated to the transaction layer
        by means of a \emph{\LINK{} data confirmation}.
    \end{description}

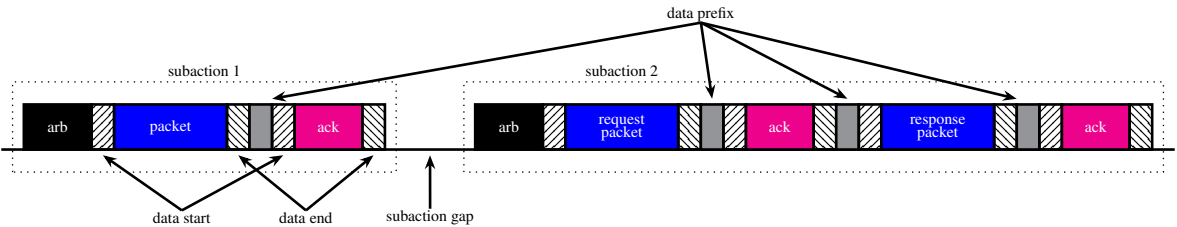
\begin{figure}[htb]
\begin{center}
\begin{tikzpicture}[line width=1pt,scale=1.5]
\draw (0,0) -- (10.4,0);

\draw[dotted,line width=.5pt ] (0.1,-0.2) rectangle (3.5,.6);
\draw (1.8,.7) node {\tiny subaction 1};

\draw[fill=black] (.2,0) rectangle (.8,.4);
\draw (.5,.2) node {\tiny \textcolor{white}{arb}};

\draw[fill=gray, pattern=north east lines] (.8,0) rectangle (1,.4);

\draw[fill=blue] (1,0) rectangle (2,.4);
\draw (1.5,.2) node {\tiny \textcolor{white}{packet}};

\draw[fill=gray,pattern=north west lines] (2,0) rectangle (2.2,.4);

\draw[fill=gray] (2.2,0) rectangle (2.4,.4);

\draw[fill=gray,pattern=north east lines] (2.4,0) rectangle (2.6,.4);

\draw[fill=magenta] (2.6,0) rectangle (3.2,.4);
\draw (2.9,.2) node {\tiny \textcolor{white}{ack}};

\draw[fill=gray, pattern=north west lines] (3.2,0) rectangle (3.4,.4);

\draw (1.6,-.6) node[inner sep=0pt] (ds) {\tiny data start};
\draw[-{Stealth[scale=0.7]}] (ds.north) -- (.9,-0.05);
\draw[-{Stealth[scale=0.7]}] (ds.north) -- (2.5,-0.05);

\draw (2.7,-.6) node[inner sep=0pt] (de) {\tiny data end};
\draw[-{Stealth[scale=0.7]}] (de.north) -- (2.1,-0.05);
\draw[-{Stealth[scale=0.7]}] (de.north) -- (3.3,-0.05);

\draw (3.8,-.6) node[inner sep=0pt] (sg) {\tiny subaction gap};
\draw[-{Stealth[scale=0.7]}] (sg.north) -- (3.8,-0.05);

\draw[dotted,line width=.5pt ] (4.1,-0.2) rectangle (10.3,.6);
\draw (5.5,.7) node {\tiny subaction 2};

\draw[fill=black] (4.2,0) rectangle (4.8,.4);
\draw (4.5,.2) node {\tiny \textcolor{white}{arb}};

\draw[fill=gray, pattern=north east lines] (4.8,0) rectangle (5,.4);

\draw[fill=blue] (5,0) rectangle (6,.4);
\draw (5.5,.2) node {\tiny \textcolor{white}{$\begin{array}{@{}c@{}}\text{request}\\[-2pt] \text{packet}\end{array}$}};

\draw[fill=gray,pattern=north west lines] (6,0) rectangle (6.2,.4);

\draw[fill=gray] (6.2,0) rectangle (6.4,.4);

\draw[fill=gray,pattern=north east lines] (6.4,0) rectangle (6.6,.4);

\draw[fill=magenta] (6.6,0) rectangle (7.2,.4);
\draw (6.9,.2) node {\tiny \textcolor{white}{ack}};

\draw[fill=gray, pattern=north west lines] (7.2,0) rectangle (7.4,.4);

\draw[fill=gray] (7.4,0) rectangle (7.6,.4);

\draw[fill=gray,pattern=north east lines] (7.6,0) rectangle (7.8,.4);

\draw[fill=blue] (7.8,0) rectangle (8.8,.4);
\draw (8.3,.2) node {\tiny \textcolor{white}{$\begin{array}{@{}c@{}}\text{response}\\[-2pt] \text{packet}\end{array}$}};

\draw[fill=gray,pattern=north west lines] (8.8,0) rectangle (9,.4);

\draw[fill=gray] (9,0) rectangle (9.2,.4);

\draw[fill=gray,pattern=north east lines] (9.2,0) rectangle (9.4,.4);

\draw[fill=magenta] (9.4,0) rectangle (10,.4);
\draw (9.7,.2) node {\tiny \textcolor{white}{ack}};

\draw[fill=gray, pattern=north west lines] (10,0) rectangle (10.2,.4);

\draw (6.2,1.2) node[inner sep=0pt] (dp) {\tiny data prefix};
\draw[-{Stealth[scale=0.7]}] (dp.south) -- (2.4,.45);
\draw[-{Stealth[scale=0.7]}] (dp.south) -- (6.3,.45);
\draw[-{Stealth[scale=0.7]}] (dp.south) -- (7.5,.45);
\draw[-{Stealth[scale=0.7]}] (dp.south) -- (9,.45);

\end{tikzpicture}
\end{center}
\caption{Subactions}\label{fig:subactions}
\end{figure}

The link layer divides the stream of data that it receives from the
physical layer into an alternating sequence of \emph{subactions} and
\emph{subaction gaps}, the latter being time intervals with a
specified minimal length during which serial cable resides in an idle state
(see Figure~\ref{fig:subactions}). A subaction either consists of a
single packet (in case of a \emph{split transaction}, see subaction 1)
or of two packets (in case of a \emph{concatenated response
  transaction}, see subaction 2).  Within each subaction, a packet is
delimited by special \emph{data start} and \emph{data end}
signals\footnote{These and other ``signals'' of the link layer
correspond to analog signals detected or emitted by the physical layer.};
the gap between two packets within a subaction must be filled with
\emph{data prefix} signals in order to distinguish these gaps from the
subaction gaps.

Before a packet can be sent, the link layer must first gain access by
issuing an arbitration procedure. Moreover, the link layer must
transform the requests of the transaction layer into a certain packet
format, computing and attaching checksums to parts of the data to be
transmitted. It also decides whether incoming packets have been
received properly by verifying the attached checksums. Every packet
that is sent by any of the nodes is received by the link layer of
every node. If a link layer determines that the packet was indeed
addressed to the node it is part of, then it forwards the contents of
the packet to the transaction layer. The link layer also handles the
sending and receiving of acknowledgements.

\item The physical connection between a node and the serial line is
  called the physical layer (the lower layer, indicated by \PHY{} in
    Figure~\ref{fig:architecture}). It listens to and puts signals
  on the serial cable, measures the lengths of the time intervals during
  which the cable resides in an idle state, and determines, together
  with the other physical layers, which node has control over the cable
  (arbitration).  It provides the following services to the link layer:

  \begin{description}
    \item[Arbitration request/confirmation:] The link layer
      instructs the physical layer to start an arbitration procedure
      by means of a \emph{\PHY{} arbitration  request}.  The result of
      this procedure (either \won{} or \lost{}) is communicated to the
      link layer by means of a \emph{\PHY{} arbitration confirmation}.
  \item[Data request/indication:] The link layer instructs the
    physical layer to put some signal on the cable by means of a
    \emph{\PHY{} data request}. The physical layer indicates to the link layer
    the detection of a signal on the cable (or information about the
    status of the cable) by means of a \emph{\PHY{} data indication}.

\item[Clock indication:] To notify the link layer that it can (and
  should) put a signal on the cable, the physical layer communicates a
  \emph{\PHY{} clock indication}.
\end{description}
\end{enumerate}

\noindent
According to \cite{IEEE-1394}, there is also a so-called \emph{node
  controller} that can influence each of the three layers. Since, in
asynchronous mode, the role of this node controller is restricted to
the ability to reset each of the three layers (force them into their
initial state), we will not consider the node controller in this paper.


\subsection{Link layer}

We proceed to describe in more detail the behaviour in asynchronous
mode of the link layer (the middle layer of the three-layered protocol),
which is responsible for the construction of packets, the transmission
of these over a serial (one-bit) line to other parties, and the
computation and verification of checksums.

We model the process behaviour of the link layer according to the
state machine depicted in \cite[Figure 6-19, Page 170]{IEEE-1394}
and the accompanying informal
explanation. The part of the state machine defining the behaviour in
asynchronous mode has eight states L$n$ ($0\leq n\leq 7$).

The link layer processes maintain a buffer (initially empty) to store a
pending request from the transaction layer.

In its initial state, the link layer can either receive a data request
from the transaction layer or a data indication from the \PHY{} layer.

At a data request, a packet is constructed from the parameters that
have been put into the buffer by the transaction layer. The link layer
process then starts a fair arbitration procedure to gain access to the
bus.  If it wins the arbitration, then the underlying
physical layer controls the cable and the link layer enters
\emph{send mode} (see below). However, it may also happen that the
physical layer indicates the arrival of data: the packet to be sent is
then stored in the buffer and the data is received first.

At a data indication, it must be checked whether the received signal
is a \ssig{} signal.  If so, this means that some other
node has control over the cable and is sending a packet; the incoming
packet must be received in \emph{receive mode}. Otherwise, the signal
(which is not a \ssig{} signal) can be ignored.

\paragraph{Send mode.}
As  soon as a node has gained control over the cable, its physical
layer starts emitting clock indications to inform the link layer that
it should send a signal. The link layer must respond to every such
clock indication and send the entire packet, one signal
at a time, delimited by a \ssig{} and an \esig{} signal. The \esig{} signal
also notifies the physical layer that the link layer is done sending
the packet; it will cease to send clock indications.  Depending on
the value of the destination field, the link layer either informs the
transaction layer that a broadcast packet was sent properly, or that it
must wait for an acknowledge packet.

The acknowledge packet must arrive within some specific amount of time: if a
subaction gap (\subactgap{} signal) occurs before an acknowledgement
with valid checksum has been received entirely (i.e., up to and including
the terminating \esig{} signal), then the link layer will act as if
the acknowledgement is missing (an acknowledge packet can
be identified by its length; it consists of one signal). When a \ssig{}
signal has been received, then the link layer expects to receive an
acknowledge signal. If the next signal is indeed a data signal, then
the link layer receives the terminating \esig{} signal, checks the
validity of the received acknowledge signal, and sends an
\emph{acknowledgement received} (\ackrec{}) to the transaction
layer. If, instead, another data signal arrives, or if there is no
terminating \esig{} signal, or if the acknowledge packet is invalid, then
the link layer sends \emph{acknowledgement missing} (\ackmiss{}) to
the transaction layer.  Both in case of failure and in case of
success, the link layer does wait for an indication of the physical
layer that a subaction gap has occurred, before
it returns to its initial state. Of course, if a subaction gap interferes
in the above described behaviour, then the link layer should
immediately send an \ackmiss{} and return to its initial state.

\paragraph{Receive mode.}
If the link layer receives a \ssig{} signal, it enters
\emph{receive mode}, expecting to see a packet being put on the
bus by some other link layer. Asynchronous packets consist of
four signals. The link layer must receive at least two
signals before it can determine whether the packet is addressed to it.

If it only receives one signal followed by a terminating \esig{}, this is
an acknowledge packet, which should be ignored: the link layer will
wait for the next subaction gap and return to the initial state.

If the second signal is indeed a  destination signal, the link layer must
check whether the incoming packet is either a packet addressed to it,
or a broadcast packet, or a packet for some other node. In the first
case, the link layer must notify the physical layer that it wants access to
the bus as soon as the packet has been received entirely, in anticipation
of sending an acknowledgement. This is done by means of an
\immediat{} arbitration request. Broadcast packets, however, are not
acknowledged; so, in the second case, no such request is needed. In the
third case, the link layer should completely ignore the packet and
return to the initial state at the next subaction gap.

The third signal is expected to be a header signal, and the fourth
signal should be a data signal. If the packet is correctly terminated
by either an \esig{} signal or a \prefix{} signal, then the packet is
forwarded to the transaction layer, either as a broadcast packet or as
a packet that was addressed to this node. In both cases, the data
checksum is verified. Observe that, in the broadcast case, a packet
with an invalid data checksum is ignored. In the other case, the
packet will have to be acknowledged, so upon winning a \emph{\PHY{}
  Arbitration confirmation}, the link layer continues in \emph{send
  acknowledgement mode}.

Any deviation of the above described procedure will cause the link
layer to ignore the packet; it will wait for a subaction gap and then
returns to the initial state. Since an \immediat{} arbitration
request may have been dispatched, a
\emph{\PHY{} Arbitration confirmation} of \won{} may still arrive.  In
such a case, the link layer is granted access to the bus, but
does not need to send an acknowledgement. Therefore, if the
destination signal indicated that the packet was meant for this node,
the arbitration confirmation must be received, and control over the
cable must be terminated immediately by sending an \esig{} signal.

\paragraph{Send acknowledgement mode.}

While the link layer is waiting for the transaction layer to respond
to a data indication with the proper acknowledgement code, it must
keep the cable busy by sending a \prefix{} signal at every clock
indication; this is to avoid the occurrence of a subaction
gap. Depending on the type of the received packet, the transaction layer
may need to issue a so-called \emph{concatenated response} (for
instance, the packet was a read request and the transaction layer
immediately wants to send the requested data to the requesting
node). By means of a data response, the transaction layer communicates
the proper acknowledgement, as well as one of the values \release{} or
\hold{}. The former means that no concatenated response is requested
and that, after sending the acknowledgement, the link layer may
release the bus and return to its initial state. The latter means
that a concatenated response is requested and that the link layer
should maintain control over the bus after sending the acknowledgement
packet by responding to clock indications with \prefix{} signals. Upon
a data request, the link layer can then go into \emph{send mode} immediately.


\subsection{Physical layer}

To simulate and analyse the interaction of the link layers of $n$
nodes, we need to model the external behaviour of underlying $n$
physical layers connected by a cable, which, together, we shall refer
to as the \emph{bus}.

The bus needs to keep track of which of the $n$ nodes have had
control over the bus during a so-called \emph{fairness interval}; to
this aim, it maintains a table of $n$ Booleans. During a fairness
interval, each node is allowed to gain control over the bus at most once,
by means of a \fair{} arbitration request. It may also access the
bus more than once as a consequence of an \immediat{} arbitration
request. As soon as the bus has been idle for some specified amount
of time and at least one link layer has got access during the
running fairness interval, an \emph{arbitration reset gap} occurs to
indicate that every node may, again, be granted access through \fair{}
arbitration. The time interval that the bus must idle before
such an arbitration reset gap may occur should be longer than
that of a subaction gap.

When the bus is in idle state and the link layer of some node requests
arbitration, the bus enters \emph{decision mode}: it checks whether
the requesting node already got access during the present fairness
interval. If not, the bus confirms the arbitration request by
indicating that the node has \won{} arbitration and evolves into a
\emph{busy} state; otherwise, the bus indicates that the arbitration is
\lost{}.

When the bus is in busy state, it records which node has control over
the bus, and which nodes have requested immediate arbitration.
In this state, the bus may still receive \fair{}
arbitration requests, but they will be confirmed by reporting that the
arbitration was \lost{}. The node that must send a response to the
packet put on the bus will issue an \immediat{} arbitration request.
No confirmation is sent, however, until the
busy node releases its control. Furthermore, as long as some link layer
still needs to send signals, the appropriate clock indications
must be generated and signals must be distributed.

In distribution mode, the bus delivers signals to all nodes except
the one that dispatched it. To obtain a realistic model, the
potential loss or corruption of signals is taken into account through a
function that assigns an error value to the checksum field of header
signals, data signals, and acknowledge signals.  Moreover, an
extra dummy value will be used to describe the situation in which
packets with a invalid length are delivered. The following
transmission errors are modelled:

\begin{itemize}
  \item If the signal is a destination signal, then this signal may
    be invalidated.  However, if this happens, the header checksum
    (which comes with the next signal) is no longer valid. The bus
    should register of which nodes invalid destinations have been
    distributed.
  \item Any signal, except for header signals having a
    corrupted checksum according to the above, may be delivered
    correctly.
  \item If the signal to be delivered is a header signal, a
    data signal or an acknowledge signal, then it may be delivered
    corrupted, or it may not be delivered at all.
  \item If the signal to be delivered is a data signal, then the
    packet may be extended by sending a dummy signal immediately after
    the data signal.
\end{itemize}

\noindent
When a signal has been distributed to every node, it is checked
whether this signal was an \esig{} signal.  If so, the
current busy node no longer requires access to the bus. It
is then checked whether some node has
requested \immediat{} arbitration. If not, a
\subactgap{} is distributed to all nodes and the bus returns to its
idle state. Otherwise, if other nodes have requested access,
control over the bus must go to one of those nodes. The bus then
sends arbitration confirmations and a clock indication to all nodes
that requested \immediat{} arbitration.

It may happen that more than one node has control over the bus.
To resolve such a conflict situation, the bus must wait for \esig{}
signals from nodes, until only one
node has access. Then, a data request is received from this node.
If it is not an \esig{} signal, the node becomes the busy one and this
signal is distributed to all other nodes. However, if
the received signal is an \esig{} signal, no node has
control over the bus anymore; a \subactgap{} signal is then distributed
to all nodes, after which the bus returns to its idle state.


\subsection{Transaction and application layers}
\label{TRANS-APPLI-NODE}

To precisely model the lower layers of IEEE~1394, it is sufficient to combine
in parallel $n$ \LINK{} processes and one \BUS{} process, which describes
$n$ \PHY{} processes and a cable. The $\mu$CRL and mCRL2 models given
in Annexes~\ref{ANNEX-MCRL} and~\ref{ANNEX-MCRL2} follow this approach
for $n=2$, with a simple \MAIN{} process gathering two link layers and a bus.

For model-checking verification (i.e., using a model checker to exhaustively
explore and analyze the reachable state space), it is desirable to describe
the upper layers as well, namely, the external behaviour of the transaction
layer and of the application running on top of it. To this aim,
M.~Sighireanu introduced in her E-LOTOS model \cite{Sighireanu-Mateescu-97-a}
two additional processes: \TRANS, which represents a transaction layer,
and {\tt Application}, which describes the application and which we note
\APPLI.

\paragraph{TRANS process.}

As mentioned in Section~\ref{ARCHITECTURE}, the transaction layer provides
read, write, and lock transactions to the application. Transactions follow
the traditional four-step connection establishment of the OSI model: request,
indication, response, and confirmation. Inside the \TRANS{} process, outgoing
requests and incoming responses are handled by two sub-processes running
in parallel and synchronized together. Both types of transactions
(concatenated and split) are dealt with. Further details can be found in
\cite[Section~7]{Sighireanu-Mateescu-97-a}.

The deadlock problem mentioned in Section~\ref{INTRO} is caused by a missing
transition in the packet transmit/receive state machine of the link layer
(precisely, in the \T{Link4BRec} sub-process of the $\mu$CRL and mCRL2 models).
To fix this bug, one option is to modify the behaviour of the link layer to
insert the missing transition, as shown in \cite{Firewire-98}. Another option
(adopted in the LOTOS and LNT models to preserve compatibility with the
$\mu$CRL and mCRL2 models) is to keep the \LINK{} process unchanged and
modify instead the \TRANS{} process by removing the transition (synchronized
with the \LINK{} process) that causes the deadlock; interestingly, the
2008 revision of IEEE~1394 also kept the link-layer state machine unchanged
(see \cite[Figure 6-21, Page 162]{IEEE-1394:2008}). Finally, to determine the
behaviour of \TRANS{}, a parameter \T{v} was added, which is equal either to
\T{ok} (deadlock-free version) or to \T{ko} (original version).

\paragraph{APPLI process.}

M.~Sighireanu designed 11~different applications, which differ by
the scenario chosen among three possibilities
(see~\cite[Section~9.2]{Sighireanu-Mateescu-97-a} for details), the maximal
number of nodes connected to the bus, and the maximal number of requests
sent to the link layer. Combined with both variants of the \TRANS{} process,
this led to 22~different \MAIN{} processes, hence 22~models to be verified.

\paragraph{NODE process.}

To factorize the vast amount of duplicated code among these 22 \MAIN{}
processes, H.~Garavel introduced a new \NODE{} process that expresses the
parallel composition of three processes: a \LINK{}, a \TRANS{}, and an
\APPLI{}. Notice that, unlike the approach of
\cite[Section~9.2]{Sighireanu-Mateescu-97-a}, the \APPLI{} process is no
longer invoked from within the \TRANS{} process.


\section{Formal models}
\label{MODELS}

In this section, we present in more detail the four formal models of the
IEEE~1394 link layer, following the chronological order of their development.


\subsection{Formal model in $\mu$CRL}
\label{TAU}

The first formal model of the link layer was written in 1997 by B.~Luttik
and circulated among the COST-247 community. It was reviewed by H.~Garavel,
J.F.~Groote, and M.~Sighireanu, who provided comments that led to improvements
and simplifications. It was published as an annex (nicely compacted using
mathematical symbols) in \cite{Luttik-97-a,Luttik-97-b} and, since then,
has remained fairly stable.
The $\mu$CRL model given in Annex~\ref{ANNEX-MCRL} is close to this
original model, with three enhancements:

\begin{itemize}
	\item It is ``machine-readable'', meaning that it can be executed by
the $\mu$CRL toolset.

	\item It uses the \B{map} keyword added in the 1997~version
of $\mu$CRL \cite{Groote-97} to declare non-constructors, whereas the
original model \cite{Luttik-97-a,Luttik-97-b} used the 1995 version
of $\mu$CRL \cite{Groote-Ponse-95}, which does not distinguish between
 constructors and non-constructors.

	\item It introduces \T{tau} internal actions in the \T{Resolve} and
\T{Distribute} sub-processes of the  \BUS{} process, in order to eliminate
two unguarded recursive calls that existed in the original model and that
the $\mu$CRL toolset cannot handle --- even if the recursion is actually bounded
by the fixed number of \LINK{} processes.
\end{itemize}

\noindent
Notice that the $\mu$CRL model is quite large (809 non-blank lines),
as the \T{Bool} and \T{Nat} types with all their basic functions must be
defined in extension. This verbosity issue was solved in the three other
formal models.


\subsection{Formal model in LOTOS}

In 1997, M.~Sighireanu wrote a LOTOS model of the IEEE~1394 link layer, based
on the draft $\mu$CRL model of B.~Luttik. The development of both models at the
same time led to clarifications, enhancements, and simplifications in each of
them. The LOTOS model aimed at using the existing CADP
toolset \cite{Fernandez-Garavel-Kerbrat-et-al-96} to perform model-checking
verification, and became an official demo example \cite{Cadp-demo-23} of CADP
in 1997. The LOTOS code was similar in essence to the $\mu$CRL code, but with
a few differences:

\begin{itemize}
	\item As mentioned in Section~\ref{TRANS-APPLI-NODE}, it introduced
\TRANS{} and \APPLI{} processes to describe the upper layers of IEEE~1394,
as well as various \MAIN{} processes specifying 22~verification scenarios.

	\item The LOTOS model was shorter because it imported predefined
libraries containing, e.g., the \T{Boolean} and \T{NaturalNumber} types.

	\item The LOTOS model uses conditional rewrite rules (e.g.,
$C_1, ..., C_n \implies L = R$) where the $\mu$CRL model needs to take
a detour via user-defined $\T{if}(C,E,E')$ functions to express
conditional equalities.

	\item The $\mu$CRL rewriter does not consider a fixed ordering of
the rewrite rules: it is the modeller's responsibility to define a confluent
term rewrite system. On the contrary, the C{\AE}SAR.ADT
compiler \cite{Garavel-89-c} for LOTOS assumes that the rewrite rules
defining each (non-constructor) function are ordered by decreasing priority;
this allows more concise definitions of equality functions (e.g., the \T{eq}
comparator for type \T{SIGNAL} has 16~rules in $\mu$CRL and 2 in LOTOS)
and other functions (e.g., \T{is\_dest}, \T{is\_header}, \T{is\_data}, and
\T{is\_ack} need 10 rules each in $\mu$CRL and 2 in LOTOS).

	\item The LOTOS model renames all local variables \T{i} to \T{j},
because the former is a reserved LOTOS keyword that denotes the internal
action (i.e., Milner's $\tau$ action). Later versions of CADP lifted this
restriction by making it possible to have LOTOS variables or
functions named \T{i}.
\end{itemize}

\noindent
This LOTOS model remained stable for many years with only, in 2005, a
simplification of the handwritten C~code used to iterate over data domains,
which was reduced from 2134 to 156 lines by factorizing similar code
fragments present in the various scenarios.

However, in 2023, H.~Garavel did a full revision of the LOTOS model, prompted by
the development of the LNT model in parallel. The volume of LOTOS code
was reduced by one third (from 2091 to 1385 lines), without loss of
functionality and still preserving strong bisimilarity. This was done
by merging the two versions of the \TRANS{} process into one parameterized
process, by merging the five versions of the \APPLI{} process into another
parameterized process, and by introducing the \NODE{} process to factorize
duplicated LOTOS code. A few other changes were made to simplify the LOTOS
code and make it closer to the $\mu$CRL code:

\begin{itemize}
	\item Like in the $\mu$CRL model, two LOTOS processes \T{Link} and
\T{Bus} have been added to serve as main entry points.

	\item The definitions of the LOTOS type \T{SIGNAL} and of its related
types have been aligned on the $\mu$CRL ones by eliminating unnecessary
auxiliary tuple types. Yet, to make the LOTOS model easier to understand,
the four overloaded constructors \T{sig} of type \T{SIGNAL} have been renamed
to \T{destsig}, \T{acksig}, \T{datasig}, and \T{headersig}, respectively
(even if LOTOS and LNT also support overloading of constructor functions).

	\item To reflect the model-checking assumptions of
\cite[Section~9.2]{Sighireanu-Mateescu-97-a}, each of the three types
\T{DATA}, \T{HEADER}, and \T{ACK} is directly defined as a singleton
(one-value) type, rather than defining it as a two-value type and later
providing ad~hoc C code that only enumerates one of these two values.
\end{itemize}


\subsection{Formal model in mCRL2}

In June 2005, the $\mu$CRL model was translated to mCRL2 by J.F.~Groote and
distributed as a demo example \cite{mCRL2-1394} in the mCRL2 toolset.

The mCRL2 spec is 60\% shorter than the $\mu$CRL one (809 non-empty lines
in $\mu$CRL vs 327 in mCRL2). Most of this reduction comes from data type
definitions, the size of which was roughly divided by 6.4 in mCRL2. This is
explained by two factors:

\begin{itemize}
	\item Like LOTOS, mCRL2 benefits from built-in data types (e.g.,
\T{Bool}, \T{Nat}, etc.), together with their basic functions, which need
not be defined in every model.

	\item Like functional languages (ML, Haskell, etc.) and E-LOTOS
\cite{ISO-15437}, mCRL2 types can be defined by their constructors.
For instance, the \T{SIGNAL} type is defined using the \B{struct} construct of
mCRL2 and the \T{BoolTABLE} type is concisely defined using the built-in
\T{List} datatype. For such types, equality functions, recognizers
(i.e., functions, such as \T{is\_dest}, that check whether an expression
matches a given constructor), and projections (i.e., functions, such as
\T{first}, \T{second}, \T{third}, and \T{fourth} for type \T{quadruple},
that extract the various arguments of a constructor) are defined automatically.
\end{itemize}

\noindent
The mCRL2 processes differ on minor points from the $\mu$CRL ones:

\begin{itemize}
	\item The syntax of the ``$\B{if } C \B{ then } A \B{ else } B$''
construct has changed: it is noted ``$C \T{ -> } A \T{ <> } B$'' in mCRL2
and ``$A \T{ <| } C \T{ |> } B$'' in $\mu$CRL.

	\item In the \LINK{} process, the $\mu$CRL definitions of the
\T{Link0} and \T{Link7} sub-processes contain summations (i.e.,
nondeterministic choices) ranging over natural numbers that are not
restricted in any way. In the mCRL2 model, these summations are bounded
by the number of \LINK{} layers.

	\item In the mCRL2 model, each \T{tau} action introduced to guard
recursion (see Section~\ref{TAU}) is replaced by an action \T{internal},
which is later abstracted from.
\end{itemize}


\subsection{Formal model in LNT}

Besides developing a complete LOTOS model and using it for model-checking
verification, M.~Sighireanu also wrote an E-LOTOS model of the IEEE~1394
link layer that was, rather than the LOTOS model itself, presented in
\cite{Sighireanu-Mateescu-97-a,Sighireanu-Mateescu-97-b,Sighireanu-Mateescu-98}.
At this time, the E-LOTOS language was still being standardized and not
finalized yet. In essence, the E-LOTOS model bears similarities with the
mCRL2 model developed later, notwithstanding the syntactic differences
between both languages.

The LNT model presented in Annex~\ref{ANNEX-LNT} does not derive from this
E-LOTOS model, as its history is distinct. In 2022, the LOTOS model (taken
in its original version) was partly translated to LNT by Oussama Oulkaid and
Marck-Edward Kemeh, as part of an exercise for master students at the
University of Grenoble. Their model was later reworked and reshaped by
H.~Garavel, in order to make it complete and strongly bisimilar to
the LOTOS one. Because it had been obtained by systematic translation,
this LNT model was very much in the same style as the $\mu$CRL, mCRL2,
and LOTOS ones: namely, data types defined as term rewrite systems, and
processes defined as state machines extended with local variables that
can be read and modified on transitions.

Therefore, H.~Garavel entirely revised this LNT model in order to obtain
a ``better'' model that would exploit the characteristic features
of LNT and demonstrate the full capabilities of this language. This revision
was achieved by progressive transformations, checking at each step that
strong bisimilarity is preserved. Concerning data specifications in the
resulting LNT model, three main remarks can be made:

\begin{itemize}
	\item The type definitions in LNT are similar (up to syntax) to
mCRL2 ones, except that equality/inequality functions must be requested
explicitly (using ``$\B{with } =$'' and ``$\B{with } <>$'' clauses) and
that functions for extracting/updating constructor arguments must also be
requested (using ``$\B{with } \T{get}$'' and ``$\B{with } \T{set}$'' clauses);
this ensures that LNT models are self-contained and not cluttered with
useless implicit functions.

	\item As regards function definitions, the LOTOS rewrite rules
ordered by decreasing priority can be systematically translated to LNT
pattern-matching \B{case} statements. However, this is not the only style
permitted by LNT, and not necessarily the most concise and readable one.
One can also define functions in a more imperative style, with the usual
programming constructs (variable assignments, \B{if}-\B{then}-\B{else},
\B{return} statements, etc.), as shown, for instance, in the various
functions manipulating values of type \T{BoolTABLE}.

	\item A salient difference between $\mu$CRL, LOTOS, and mCRL2, on
the one hand, and LNT, on the other hand, concerns partial functions,
i.e., functions that are not defined over the entire
domain of their arguments (e.g., function \T{get} for the \T{BoolTABLE}
type or functions \T{getdest}, \T{getdcrc}, \T{getdata}, \T{gethead},
\T{getadd}, and \T{corrupt} for the \T{SIGNAL} type).
In $\mu$CRL, LOTOS, and mCRL2, partial definition is implicit, in
the sense that some equations are not given, e.g., there is no equation
to define ``\T{get ($n$, empty)}''. The LOTOS model of Annex~\ref{ANNEX-LOTOS}
contains comments to warn about partial definitions, but this is left to the
good will of the specifier.

	In LNT, the situation is different: any partial function triggers
(based on control- and data-flow analysis) an error, which the specifier
is expected to correct, either by properly dealing with the overlooked cases,
or by explicitly inserting a ``$\B{raise } E$'' statement at each point
where the function might terminate without returning a result --- $E$ being
either an event declared as an exception that the function can raise,
or the predefined event \T{UNEXPECTED} denoting an exception that cannot
be caught and triggers a run-time error.
\end{itemize}

\noindent
Concerning processes, the following five transformations have been
repeatedly applied until an idiomatic LNT model was obtained:

\begin{itemize}
	\item The guarded commands
``$\T{[} C \T{]} \rightarrow A \T{ [] } \T{[} \T{not}(C) \T{]} \rightarrow B\,$''
present in the LOTOS model have been translated to
``$\B{if } C \B{ then } A \B{ else } B \B{ end if\,}$'' statements of LNT.
The \B{then} and \B{else} branches have been permuted, negating the
Boolean condition $C$, when $B$ was much shorter than $A$.
Also, nested \B{if} statements have been flattened whenever possible by
using the (Ada-like) \B{elsif} clause of LNT.

	\item When this was convenient, calls to recursive processes have
been replaced by the \B{loop} statements of LNT, possibly with a \B{break}
statement to exit the loop. For instance, the \T{Link3}, \T{Link5}, and
\T{Link7} processes of the $\mu$CRL, mCRL2, and LOTOS models have been
replaced,  in the LNT model, by \B{loop} statements. Indeed, in $\mu$CRL,
mCRL2, and LOTOS, (finite or infinite) iteration must always be expressed
using recursion, with two main drawbacks:
(i) the mandatory use of recursion obfuscates the
flow of control by requiring the definition of auxiliary recursive processes
and ``goto-like'' calls to these processes; (ii) it also obfuscates the
flow of data by requiring, for such processes, as many parameters as there
are live variables at the point where these processes are called.
Using iteration rather than recursion often leads to simpler, more
readable models.

	\item In some cases, finite loops can be further simplified by
turning them into \B{while} or \B{for} loops. For instance, the sub-process
\T{Resolve2} of the $\mu$CRL, mCRL2, and LOTOS models can be rephrased as
a \B{while} loop, whereas the sub-processes \T{Resolve}, \T{SubactionGap},
and \T{Distribute} can be described using \B{for} loops, hereby getting rid
of the extra parameters that store the loop variables. Notice that
such iterative behaviour was quite clear from the textual description of
these processes in \cite{Luttik-97-a}, but only LNT enables one to express
it in natural way.

	\item Processes that are called only once (especially after recursion
has been replaced by iteration) should be expanded in-line at the point where
they are called. Doing so, the control flow becomes more readable (as each
process call is similar to a ``goto'') and many process parameters are
eliminated. M.~Sighireanu applied this idea when designing her E-LOTOS
model: the two $\mu$CRL sub-processes \T{DecideIdle} and \T{Link1}
were expanded in-line \cite[footnotes 7 and 8]{Sighireanu-Mateescu-97-a}.
In the LNT model of Annex~\ref{ANNEX-LNT}, this idea was pushed beyond
by also eliminating the sub-processes \T{Link3}, \T{Link3RA}, \T{Link3RE},
\T{Link4DH}, \T{Link4RH}, \T{Link4RD}, \T{Link4RE}, \T{Link4BRec},
\T{Link4DRec}, \T{Link5}, \T{Link6}, \T{Link7}, \T{Resolve}, and
\T{Resolve2}. The sub-process \T{Link4}, although called only once, was
not expanded in-line, because it is so large that its expansion would have
increased the nesting depth too much. Also, a new \T{Link2} sub-process was
added to factorize both sub-processes \T{Link2req} and \T{Link2resp} in a
single one. As a result, the \LINK{} process has only 6 (mutually recursive)
processes in the LNT model, instead of 19 in the other models --- maintaining
an exact correspondence with the 8~states describing the asynchronous mode
\cite[Figure 6-19, Page 170]{IEEE-1394} was not considered a requirement
for the LNT model.

	\item Since the in-line expansion of processes often creates variables
with nested scopes, three additional transformations may be suitable to keep
the LNT model simple:
\begin{itemize}
\item
merging different variables that have the same type and are never used
simultaneously, so as to decrease the number of variables.
\item
enlarging the scope of nested variables by moving their declarations
upward, so has to reduce the nesting depth of variable scopes;
\item
renaming nested variables declared in the scope of another variable having
the same name; for instance, after successively expanding the sub-process
\T{Link7} in \T{Link6}, \T{Link6} in \T{Link5}, and \T{Link5} in \T{Link4DRec},
the \T{d} variable of \T{Link7} arrives in the scope of the \T{d} variable
of \T{Link4DRec}; even if the innermost variable hides the outermost one
in LNT (as in Algol-60), it may be suitable to give these variables different
names to avoid confusion.
\end{itemize}
These transformations sometimes conflict with each other, and their judicious
application cannot be governed by strict laws: it is rather a matter of taste
and circumstances.
\end{itemize}


\section{Verification}
\label{VERIFICATION}

The four formal models of the IEEE~1934 link layer have been checked by
their respective compilers: the $\mu$CRL toolset, the mCRL2 toolset, and,
for the LOTOS and LNT models, the CADP toolset.

The five correctness properties stated by B.~Luttik
\cite[Section~4]{Luttik-97-a} have been formulated in the ACTL temporal
logic~\cite{DeNicola-Vaandrager-90} by R.~Mateescu and M.~Sighireanu
\cite[Section~10]{Sighireanu-Mateescu-97-a}. Using the
XTL~\cite{Mateescu-Garavel-98} model checker of CADP, these formulas have
been checked on 16 out of 22 variants of the LOTOS model (totalling 80
model-checking jobs), the domains of the types \T{ACK}, \T{DATA}, and
\T{HEADER} being limited to a single value. All the properties
hold, except the first property (deadlock freeness), which is violated on
the ``original'' models when the application layer executes its most
complex scenarios.

The LNT model has been verified in two ways, using both model checking
and equivalence checking. On the one hand, the ACTL formulas evaluate
identically on the 16 variants of the LNT model. On the other hand,
the labelled transition systems generated from 20 out of 22 variants
of the LNT model are strongly bisimilar to those generated from
the same variants of the LOTOS model. The labelled transition systems
of the two remaining variants are too large for being generated directly,
and would certainly benefit from compositional verification techniques
\cite{Garavel-Lang-Mounier-18}. In 14~cases out of 20, the labelled transition
systems generated from LOTOS and LNT have the same size, whereas in 6~cases,
those generated from LNT are slightly larger (+0.46\% states, +0.43\%
transitions). Using version 2024-a ``Eindhoven'' of the CADP toolbox,
these verifications were performed in less than 8 minutes on a Dell
Latitude 5580 (Intel Core i5-7200U processor, 16~GB RAM) running Linux.


\section{Conclusion}
\label{CONCLUSION}

Revisiting the IEEE~1394 link layer problem, a true success story of
formal methods, we presented and discussed four models written in $\mu$CRL,
mCRL2, LOTOS, and LNT --- the LOTOS model (revised in 2023) and the LNT model
being novel contributions. In this respect, the present paper is a tentative
``Rosetta stone'' for comparing various modelling languages dedicated to
communication protocols and concurrent systems. In a nutshell, our main
findings are as follows:

\begin{itemize}
	\item It appears that the three languages $\mu$CRL, mCRL2, and LOTOS
are quite close, except that data type specifications are more concise in
the latter two languages. Each of these three languages contains two
separate sub-languages: one for specifying data types (using algebraic
specifications or term rewrite rules), and another one for concurrent
processes.

These sub-languages sometimes use distinct symbols to express
the same concept (e.g., \B{if}-\B{then}-\B{else} being noted differently
in the data and process parts) and sometimes give the same
symbol totally different meanings, e.g., in $\mu$CRL and mCRL2, the ``$+$''
operator (which denotes addition in the data part and nondeterministic choice
in the  process part), the ``$||$'' operator (which denotes logical disjunction
in the data part and parallel composition in the process part), or closing
parentheses (which denote the end of expressions in the data part and the
end of a choice, a sequential composition, etc. in the process part).

On the contrary, LNT is a unified language, without separate sub-languages:
LNT functions and LNT processes are defined using the same notations
(``;'' for sequential composition, \B{if}-\B{then}-\B{else} for conditionals,
etc.), and LNT avoids, as much as possible, ``overloaded'' symbols.

	\item Although it has been argued that LOTOS supports very diverse
 ``specification styles'' \cite{Vissers-Scollo-vanSinderen-Brinksma-91},
most LOTOS, $\mu$CRL, and mCRL2 models consist of a set of concurrent
processes, each of which being specified using guarded commands and
terminal recursion. Such a style is convenient for describing automata
extended with state variables, but leads to models that are difficult to
maintain when specifications evolve frequently, and does not scale well
when automata complexity increases, resulting in large, poorly structured
state machines scattered with ``goto-like'' transitions.

	In addition to supporting guarded commands and terminal recursion,
LNT provides alternative specification styles suitable for the description
of complex systems. In particular, LNT offers the classical primitives of
structured programming, properly bracketed with an Ada-like syntax, which
make large models easier to read and reduce the need for drawing state
machines on paper.
\end{itemize}

\noindent
To some extent, there is here a debate around the concept of
minimality and how it should be interpreted. On the one hand, LOTOS,
$\mu$CRL, and mCRL2 try to be minimal in the size of the
language\footnote{The $\mu$ letter (which stands for ``micro'') in
  $\mu$CRL indeed expresses such a desire for minimality.}, the number
of syntactic constructs, and the number of semantic rules. An explicit
concern for $\mu$CRL and mCRL2 has been to ensure that the semantics
are as simple and elegant as possible, only including constructs in
the language if they are needed for expressiveness; ease of modelling
has been less of a concern so far.  LNT also tries to be minimal,
e.g., by unifying the sub-languages for functions and processes, the
former being included in the latter, but it can be rightly argued that
LNT is richer than the three other languages and requires more complex
compilers that implement involved control- and data-flow analyses.

Perhaps the proper concept of minimality is not so much about the size of
a language or of its compiler, but about the effort needed to learn the
language, the time needed to write correct models, and the difficulty
of understanding such models for engineers who do not have a strong
background in formal methods. We hope that the present study will usefully
contribute to this debate.


\subsection*{Acknowledgements}

We are grateful to all those who contributed to the design and verification
of the four formal models of the IEEE~1394 link layer, namely:
Jan Friso Groote, Marck-Edward Kemeh, Radu Mateescu, Laurent Mounier,
Oussama Oulkaid, Charles Pecheur, Judi Romijn, Mihaela Sighireanu,
and Bruno Vivien. We also thank the anonymous reviewers for their
constructive remarks.


\bibliographystyle{eptcs}
\bibliography{main}


\appendix

\newcommand{\SPACING}{\vspace{2ex}}


\section{Formal model in $\mu$CRL}
\label{ANNEX-MCRL}

\lstset{
  language=mCRL,
  basicstyle=\ttfamily\small,
  columns=fullflexible, 
  commentstyle=\rmfamily\itshape,
  keywordstyle=\rmfamily\bfseries,
  xleftmargin=0pt
}


\subsection{Types and functions in $\mu$CRL}

\begin{lstlisting}

% Boolean type

sort Bool
func
  T,F: -> Bool

map
  eq: Bool#Bool -> Bool
var
  b: Bool
rew
  eq(T,b)=b
  eq(b,T)=b
  eq(b,F)=not(b)
  eq(F,b)=not(b)

map
  and: Bool#Bool -> Bool
var
  b: Bool
rew
  and(T,b)=b
  and(b,T)=b
  and(b,F)=F
  and(F,b)=F

map
  or: Bool#Bool -> Bool
var
  b: Bool
rew
  or(T,b)=T
  or(b,T)=T
  or(b,F)=b
  or(F,b)=b

map
  not: Bool           -> Bool
  if:  Bool#Bool#Bool -> Bool
var
  b1,b2: Bool
rew
  not(F)=T
  not(T)=F
  if(T,b1,b2)=b1
  if(F,b1,b2)=b2

% Natural number type

sort NAT
func
  0,1,2:         -> NAT
%  0,1,2,3,4,5,6,7,8,9:         -> NAT
map  succ:  NAT -> NAT

map
  eq: NAT#NAT -> Bool
var
  n,m: NAT
rew
  1=succ(0)
  2=succ(1)
  eq(0,0)=T
  eq(succ(n),0)=F
  eq(0,succ(n))=F
  eq(succ(n),succ(m))=eq(n,m)

map
  lt: NAT#NAT -> Bool
var
  n,m: NAT
rew
  lt(0,0)=F
  lt(succ(n),0)=F
  lt(0,succ(n))=T
  lt(succ(n),succ(m))=lt(n,m)

% Data/Control/Acknowledge elemens and their CRC computation

sort CHECK
func
  bottom,check:   -> CHECK
map
  eq: CHECK#CHECK -> Bool
rew
  eq(bottom,bottom)=T
  eq(check,check)=T
  eq(check,bottom)=F
  eq(bottom,check)=F

sort DATA
func
  d1,d2:      -> DATA
map
  crc:   DATA -> CHECK
  eq:  DATA#DATA -> Bool
rew
  crc(d1)=check
  crc(d2)=check
  eq(d1,d1)=T
  eq(d1,d2)=F
  eq(d2,d1)=F
  eq(d2,d2)=T

sort HEADER
func
  h1,h2:       -> HEADER
map
  crc:  HEADER -> CHECK
  eq: HEADER # HEADER -> Bool
rew
  crc(h1)=check
  crc(h2)=check
  eq(h1,h1)=T
  eq(h1,h2)=F
  eq(h2,h1)=F
  eq(h2,h2)=T

sort ACK
func
  a1,a2:     -> ACK
map
  crc:   ACK -> CHECK
  eq : ACK # ACK -> Bool
rew
  crc(a1)=check
  crc(a2)=check
  eq(a1,a1)=T
  eq(a1,a2)=F
  eq(a2,a1)=F
  eq(a2,a2)=T

sort SIGNAL
func
  sig: NAT          -> SIGNAL
  sig: HEADER#CHECK -> SIGNAL
  sig: DATA#CHECK   -> SIGNAL
  sig: ACK#CHECK    -> SIGNAL

  Start,End:        -> SIGNAL
  Prefix,subactgap: -> SIGNAL
  dhead,Dummy:      -> SIGNAL
map
  is_start,is_end:    SIGNAL -> Bool
  is_prefix,is_sagap: SIGNAL -> Bool
  is_dummy,is_dhead:  SIGNAL -> Bool
  eq:                 SIGNAL#SIGNAL -> Bool
var
  n,n' : NAT
  h,h' : HEADER
  d,d' : DATA
  a,a' : ACK
  c,c' : CHECK
  s : SIGNAL
rew
  is_start(Start)=T
  is_start(End)=F
  is_start(Prefix)=F
  is_start(subactgap)=F
  is_start(dhead)=F
  is_start(Dummy)=F
  is_start(sig(n))=F
  is_start(sig(h,c))=F
  is_start(sig(d,c))=F
  is_start(sig(a,c))=F
  eq(Start,s)=is_start(s)
  eq(s,Start)=is_start(s)

  is_end(End)=T
  is_end(Start)=F
  is_end(Prefix)=F
  is_end(subactgap)=F
  is_end(dhead)=F
  is_end(Dummy)=F
  is_end(sig(n))=F
  is_end(sig(h,c))=F
  is_end(sig(d,c))=F
  is_end(sig(a,c))=F
  eq(End,s)=is_end(s)
  eq(s,End)=is_end(s)

  is_prefix(Prefix)=T
  is_prefix(Start)=F
  is_prefix(End)=F
  is_prefix(subactgap)=F
  is_prefix(dhead)=F
  is_prefix(Dummy)=F
  is_prefix(sig(n))=F
  is_prefix(sig(h,c))=F
  is_prefix(sig(d,c))=F
  is_prefix(sig(a,c))=F
  eq(Prefix,s)=is_prefix(s)
  eq(s,Prefix)=is_prefix(s)

  is_sagap(subactgap)=T
  is_sagap(Start)=F
  is_sagap(End)=F
  is_sagap(Prefix)=F
  is_sagap(dhead)=F
  is_sagap(Dummy)=F
  is_sagap(sig(n))=F
  is_sagap(sig(h,c))=F
  is_sagap(sig(d,c))=F
  is_sagap(sig(a,c))=F
  eq(subactgap,s)=is_sagap(s)
  eq(s,subactgap)=is_sagap(s)

  is_dhead(subactgap)=F
  is_dhead(Start)=F
  is_dhead(End)=F
  is_dhead(Prefix)=F
  is_dhead(dhead)=T
  is_dhead(Dummy)=F
  is_dhead(sig(n))=F
  is_dhead(sig(h,c))=F
  is_dhead(sig(d,c))=F
  is_dhead(sig(a,c))=F
  eq(dhead,s)=is_dhead(s)
  eq(s,dhead)=is_dhead(s)

  is_dummy(subactgap)=F
  is_dummy(Start)=F
  is_dummy(End)=F
  is_dummy(Prefix)=F
  is_dummy(dhead)=F
  is_dummy(Dummy)=T
  is_dummy(sig(n))=F
  is_dummy(sig(h,c))=F
  is_dummy(sig(d,c))=F
  is_dummy(sig(a,c))=F
  eq(Dummy,s)=is_dummy(s)
  eq(s,Dummy)=is_dummy(s)

  eq(sig(n),sig(n'))=eq(n,n')
  eq(sig(n),sig(h,c))=F
  eq(sig(n),sig(d,c))=F
  eq(sig(n),sig(a,c))=F
  eq(sig(h,c),sig(n'))=F
  eq(sig(h,c),sig(h',c'))=and(eq(h,h'),eq(c,c'))
  eq(sig(h,c),sig(d,c'))=F
  eq(sig(h,c),sig(a,c'))=F
  eq(sig(d,c),sig(n))=F
  eq(sig(d,c),sig(h,c'))=F
  eq(sig(d,c),sig(d',c'))=and(eq(d,d'),eq(c,c'))
  eq(sig(d,c),sig(a,c'))=F
  eq(sig(a,c),sig(n))=F
  eq(sig(a,c),sig(h,c'))=F
  eq(sig(a,c),sig(d,c'))=F
  eq(sig(a,c),sig(a',c'))=and(eq(a,a'),eq(c,c'))

map
  is_dest,is_header: SIGNAL     -> Bool
  is_data,is_ack:    SIGNAL     -> Bool
var
  n : NAT
  h : HEADER
  d : DATA
  a : ACK
  c : CHECK
rew
  is_dest(sig(n))=T
  is_dest(sig(h,c))=F
  is_dest(sig(d,c))=F
  is_dest(sig(a,c))=F
  is_dest(Start)=F
  is_dest(End)=F
  is_dest(Prefix)=F
  is_dest(subactgap)=F
  is_dest(dhead)=F
  is_dest(Dummy)=F

  is_header(sig(h,c))=T
  is_header(sig(n))=F
  is_header(sig(d,c))=F
  is_header(sig(a,c))=F
  is_header(Start)=F
  is_header(End)=F
  is_header(Prefix)=F
  is_header(subactgap)=F
  is_header(dhead)=F
  is_header(Dummy)=F

  is_data(sig(d,c))=T
  is_data(sig(n))=F
  is_data(sig(h,c))=F
  is_data(sig(a,c))=F
  is_data(Start)=F
  is_data(End)=F
  is_data(Prefix)=F
  is_data(subactgap)=F
  is_data(dhead)=F
  is_data(Dummy)=F

  is_ack(sig(a,c))=T
  is_ack(sig(n))=F
  is_ack(sig(h,c))=F
  is_ack(sig(d,c))=F
  is_ack(Start)=F
  is_ack(End)=F
  is_ack(Prefix)=F
  is_ack(subactgap)=F
  is_ack(dhead)=F
  is_ack(Dummy)=F

map
  is_physig,is_terminator: SIGNAL -> Bool
var
  s : SIGNAL
rew
  is_physig(s)=or(is_start(s),or(is_end(s),or(is_prefix(s),is_sagap(s))))
  is_terminator(s)=or(is_end(s),is_prefix(s))

map
  is_hda: SIGNAL -> Bool
var
  s : SIGNAL
rew
  is_hda(s)=or(is_header(s),or(is_data(s),is_ack(s)))

map
  valid_hpart, valid_ack: SIGNAL -> Bool
var
  n : NAT
  h : HEADER
  d : DATA
  a : ACK
  c : CHECK
rew
  valid_ack(sig(a,c))=eq(c,check)
  valid_ack(sig(h,c))=F
  valid_ack(sig(d,c))=F
  valid_ack(sig(n))=F
  valid_ack(Start)=F
  valid_ack(End)=F
  valid_ack(Prefix)=F
  valid_ack(subactgap)=F
  valid_ack(Dummy)=F
  valid_ack(dhead)=F

  valid_hpart(sig(h,c))=eq(c,check)
  valid_hpart(sig(n))=F
  valid_hpart(sig(d,c))=F
  valid_hpart(sig(a,c))=F
  valid_hpart(Start)=F
  valid_hpart(End)=F
  valid_hpart(Prefix)=F
  valid_hpart(subactgap)=F
  valid_hpart(Dummy)=F
  valid_hpart(dhead)=F

map
  getdest:     SIGNAL            -> NAT
  getdcrc:     SIGNAL            -> CHECK
  getdata:     SIGNAL            -> DATA
  gethead:     SIGNAL            -> HEADER
  getack:      SIGNAL            -> ACK
  corrupt:     SIGNAL            -> SIGNAL
var
  n   : NAT
  h   : HEADER
  d   : DATA
  a   : ACK
  c   : CHECK
rew
  getdest(sig(n))    = n
  gethead(sig(h,c))  = h
  getdcrc(sig(d,c))  = c
  getdata(sig(d,c))  = d
  getack (sig(a,c))  = a

  corrupt(sig(h,c)) = sig(h,bottom)
  corrupt(sig(d,c)) = sig(d,bottom)
  corrupt(sig(a,c)) = sig(a,bottom)

sort SIG_TUPLE
func
  quadruple: SIGNAL#SIGNAL#SIGNAL#SIGNAL -> SIG_TUPLE
  void:                                  -> SIG_TUPLE
map
  first,second,third,fourth:  SIG_TUPLE  -> SIGNAL
  is_void: SIG_TUPLE -> Bool
var
  x1,x2,x3,x4: SIGNAL
rew
  first(quadruple(x1,x2,x3,x4))=x1
  second(quadruple(x1,x2,x3,x4))=x2
  third(quadruple(x1,x2,x3,x4))=x3
  fourth(quadruple(x1,x2,x3,x4))=x4

  is_void(void)=T
  is_void(quadruple(x1,x2,x3,x4))=F

sort PAR
func
  fair,immediate:       -> PAR
map
  eq: PAR#PAR -> Bool
rew
  eq(fair,fair)=T
  eq(immediate,immediate)=T
  eq(fair,immediate)=F
  eq(immediate,fair)=F

sort PAC
func
  won,lost:               -> PAC
map
  eq: PAC#PAC -> Bool
rew
  eq(won,won)=T
  eq(lost,lost)=T
  eq(won,lost)=F
  eq(lost,won)=F

sort LDC
func
  ackrec:            ACK -> LDC
  ackmiss,broadsent:     -> LDC

sort LDI
func
  good,broadrec: HEADER#DATA -> LDI
  dcrc_err:      HEADER      -> LDI

sort BOC
func
  release,hold: -> BOC
map
  eq:   BOC#BOC -> Bool
rew
  eq(release,release)=T
  eq(hold,hold)=T
  eq(release,hold)=F
  eq(hold,release)=F

\end{lstlisting}
\SPACING


\subsection{The LINK process in $\mu$CRL}

\begin{lstlisting}
act
  LDreq:  NAT#NAT#HEADER#DATA
  LDcon:  NAT#LDC
  LDind:  NAT#LDI
  LDres:  NAT#ACK#BOC

  sPDreq,rPDind: NAT#SIGNAL
  sPAreq: NAT#PAR
  rPAcon: NAT#PAC
  rPCind: NAT

proc

LINK(n:NAT,i:NAT)=
   ( Link0(n,i,void) )

Link0(n:NAT,id:NAT,buffer:SIG_TUPLE)=
(
    sum(dest:NAT,
      sum(h:HEADER,
        sum(d:DATA,
          LDreq(id,dest,h,d).
            Link0(n,id,quadruple(dhead,
                                 sig(dest),
                                 sig(h,crc(h)),
                                 sig(d,crc(d))))
        )
      )
    )
  <| is_void(buffer) |>
    sPAreq(id,fair).Link1(n,id,buffer)
)
+
  sum(p:SIGNAL,
    rPDind(id,p).
      ( Link4(n,id,buffer) <| is_start(p) |> Link0(n,id,buffer) )
  )

Link1(n:NAT,id:NAT,p:SIG_TUPLE)=
  rPAcon(id,won).Link2req(n,id,p)
+
  rPAcon(id,lost).Link0(n,id,p)

Link2req(n:NAT,id:NAT,p:SIG_TUPLE)=
   ( rPCind(id).sPDreq(id,Start).
     rPCind(id).sPDreq(id,first(p)).
     rPCind(id).sPDreq(id,second(p)) ) .
   ( rPCind(id).sPDreq(id,third(p)).
     rPCind(id).sPDreq(id,fourth(p)).
     rPCind(id).sPDreq(id,End) ).
    (
      LDcon(id,broadsent).Link0(n,id,void)
     <| eq(getdest(second(p)),n) |>
      Link3(n,id,void)
    )

Link3(n:NAT,id:NAT,buffer:SIG_TUPLE)=
  sum(p:SIGNAL,
    rPDind(id,p).
      (
        Link3(n,id,buffer)
      <| is_prefix(p) |>
        (
          Link3RA(n,id,buffer)
        <| is_start(p) |>
          (
            LDcon(id,ackmiss).Link0(n,id,buffer)
          <| is_sagap(p) |>
            LDcon(id,ackmiss).LinkWSA(n,id,buffer,n)
          )
        )
      )
  )

Link3RA(n:NAT,id:NAT,buffer:SIG_TUPLE)=
  sum(a:SIGNAL,
    rPDind(id,a).
      (
        (
          LDcon(id,ackmiss).Link0(n,id,buffer)
        <| is_sagap(a) |>
          LDcon(id,ackmiss).LinkWSA(n,id,buffer,n)
        )
      <| is_physig(a) |>
        Link3RE(n,id,buffer,a)
      )
  )

Link3RE(n:NAT,id:NAT,buffer:SIG_TUPLE,a:SIGNAL)=
  sum(e:SIGNAL,
    rPDind(id,e).
      (
        LDcon(id,ackrec(getack(a))).LinkWSA(n,id,buffer,n)
      <| and(valid_ack(a),is_terminator(e)) |>
        (
          LDcon(id,ackmiss).Link0(n,id,buffer)
        <| is_sagap(e) |>
          LDcon(id,ackmiss).LinkWSA(n,id,buffer,n)
        )
      )
  )

Link4(n:NAT,id:NAT,buffer:SIG_TUPLE)=
  sum(dh:SIGNAL,
    rPDind(id,dh).
      (
        (
          Link0(n,id,buffer)
        <| is_sagap(dh) |>
          LinkWSA(n,id,buffer,n)
        )
      <| is_physig(dh) |>
        Link4DH(n,id,buffer)
      )
  )

Link4DH(n:NAT,id:NAT,buffer:SIG_TUPLE)=
  sum(dest:SIGNAL,
    rPDind(id,dest).
      (
        (
          sPAreq(id,immediate).Link4RH(n,id,buffer,id)
        <| eq(getdest(dest),id) |>
          (
            Link4RH(n,id,buffer,n)
          <| eq(getdest(dest),n) |>
            LinkWSA(n,id,buffer,n)
          )
        )
      <| is_dest(dest) |>
        (
          Link0(n,id,buffer)
        <| is_sagap(dest) |>
          LinkWSA(n,id,buffer,n)
        )
      )
  )

Link4RH(n:NAT,id:NAT,buffer:SIG_TUPLE,dest:NAT)=
  sum(h:SIGNAL,
    rPDind(id,h).
      (
        Link4RD(n,id,buffer,dest,h)
      <| valid_hpart(h) |>
        LinkWSA(n,id,buffer,dest)
      )
  )

Link4RD(n:NAT,id:NAT,buffer:SIG_TUPLE,dest:NAT,h:SIGNAL)=
  sum(d:SIGNAL,
    rPDind(id,d).
      (
        Link4RE(n,id,buffer,dest,h,d)
      <| is_data(d) |>
        LinkWSA(n,id,buffer,dest)
      )
  )

Link4RE(n:NAT,id:NAT,buffer:SIG_TUPLE,dest:NAT,h:SIGNAL,d:SIGNAL)=
  sum(e:SIGNAL,
    rPDind(id,e).
      (
        (
          Link4DRec(n,id,buffer,h,d)
        <| eq(dest,id) |>
          Link4BRec(n,id,buffer,h,d)
        )
      <| is_terminator(e) |>
        LinkWSA(n,id,buffer,dest)
      )
  )

Link4DRec(n:NAT,id:NAT,buffer:SIG_TUPLE,h:SIGNAL,d:SIGNAL)=
  LDind(id,good(gethead(h),getdata(d))).rPAcon(id,won).Link5(n,id,buffer)
<| eq(getdcrc(d),check) |>
  LDind(id,dcrc_err(gethead(h))).rPAcon(id,won).Link5(n,id,buffer)

Link4BRec(n:NAT,id:NAT,buffer:SIG_TUPLE,h:SIGNAL,d:SIGNAL)=
  LDind(id,broadrec(gethead(h),getdata(d))).Link0(n,id,buffer)
<| eq(getdcrc(d),check) |>
  Link0(n,id,buffer)

Link5(n:NAT,id:NAT,buffer:SIG_TUPLE)=
  sum(a:ACK,
    sum(b:BOC,
      LDres(id,a,b).Link6(n,id,buffer,sig(a,crc(a)),b)
    )
  )
+
  rPCind(id).sPDreq(id,Prefix).Link5(n,id,buffer)

Link6(n:NAT,id:NAT,buffer:SIG_TUPLE,p:SIGNAL,b:BOC)=
  ( rPCind(id).sPDreq(id,Start).rPCind(id).sPDreq(id,p) ) .
      ( rPCind(id).
        (
          sPDreq(id,End).Link0(n,id,buffer)
        <| eq(b,release) |>
          sPDreq(id,Prefix).Link7(n,id,buffer)
        )
      )

Link7(n:NAT,id:NAT,buffer:SIG_TUPLE)=
  rPCind(id).sPDreq(id,Prefix).Link7(n,id,buffer)
+
  sum(dest:NAT,
    sum(h:HEADER,
      sum(d:DATA,
        LDreq(id,dest,h,d).
          Link2resp(n,id,buffer,quadruple(dhead,
                                          sig(dest),
                                          sig(h,crc(h)),
                                          sig(d,crc(d))))
      )
    )
  )

Link2resp(n:NAT,id:NAT,buffer:SIG_TUPLE,p:SIG_TUPLE)=
  ( rPCind(id).sPDreq(id,Start).
    rPCind(id).sPDreq(id,first(p)).
    rPCind(id).sPDreq(id,second(p)) ).
    ( rPCind(id).sPDreq(id,third(p)).
      rPCind(id).sPDreq(id,fourth(p)).
      rPCind(id).sPDreq(id,End)).
        (  LDcon(id,broadsent).Link0(n,id,buffer)
             <| eq(getdest(second(p)),n) |>
               Link3(n,id,buffer)
        )

LinkWSA(n:NAT,id:NAT,buffer:SIG_TUPLE,dest:NAT)=
  sum(p:SIGNAL,
    rPDind(id,p).
      (
        Link0(n,id,buffer)
      <| is_sagap(p) |>
        LinkWSA(n,id,buffer,dest)
      )
  )
+
  (
    rPAcon(id,won).rPCind(id).sPDreq(id,End).Link0(n,id,buffer)
  <| eq(dest,id) |>
    delta
  )

\end{lstlisting}
\SPACING


\subsection{The BUS process in $\mu$CRL}

\begin{lstlisting}

sort BoolTABLE
func
  empty:                             -> BoolTABLE
  btable:   NAT#Bool#BoolTABLE       -> BoolTABLE

map
  inita:     NAT                      -> BoolTABLE
  invert:   NAT#BoolTABLE            -> BoolTABLE
  get:      NAT#BoolTABLE            -> Bool
  if:       Bool#BoolTABLE#BoolTABLE -> BoolTABLE
  eq:BoolTABLE#BoolTABLE->Bool
var
  n,m     : NAT
  b       : Bool
bt1,bt2 : BoolTABLE
rew
  eq(bt1, bt1)=T
  inita(0)=empty
  inita(succ(n))=btable(n,F,inita(n))

  invert(n,empty)=empty
  invert(n,btable(m,b,bt1))=
    if(eq(n,m),
      btable(m,not(b),bt1),
      btable(m,b,invert(n,bt1))
    )
  get(n,btable(m,b,bt1))=if(eq(n,m),b,get(n,bt1))
  get(n,empty)=F
  if(T,bt1,bt2)=bt1
  if(F,bt1,bt2)=bt2

map
  zero,one,more: BoolTABLE -> Bool
var
  n  : NAT
  bt : BoolTABLE
rew
  zero(empty)=T
  zero(btable(n,T,bt))=F
  zero(btable(n,F,bt))=zero(bt)
  one(empty)=F
  one(btable(n,T,bt))=zero(bt)
  one(btable(n,F,bt))=one(bt)
  more(bt)=and(not(zero(bt)),not(one(bt)))

act
  rPAreq: NAT#PAR
  rPDreq,sPDind: NAT#SIGNAL
  sPAcon: NAT#PAC
  sPCind: NAT
  arbresgap
  losesignal

proc

BUS(n:NAT)=
  BusIdle(n, inita(n))

BusIdle(n:NAT,t:BoolTABLE)=
  sum(id:NAT,
    sum(astat:PAR,
      rPAreq(id,astat).DecideIdle(n,t,id,astat)))
+
  arbresgap.BusIdle(n,inita(n)) <| not(zero(t)) |> delta

DecideIdle(n:NAT,t:BoolTABLE,id:NAT,astat:PAR)=
  ( sPAcon(id,won).BusBusy(n,invert(id,t),inita(n),inita(n),id) )
  <| not(get(id,t)) |>
  ( sPAcon(id,lost).BusIdle(n,t) )

BusBusy(n:NAT,
        t:BoolTABLE,
        next:BoolTABLE,
        destfault:BoolTABLE,
        busy:NAT)=
(
  (
    sPCind(busy).
      sum(p:SIGNAL,
        rPDreq(busy,p).Distribute(n,t,next,destfault,busy,p,0)
      )
  )
  <| lt(busy,n) |>
  (
      SubactionGap(n,t,0)
    <| zero(next) |>
      Resolve(n,t,next,0)
  )
)
+
  sum(j:NAT,
    rPAreq(j,fair).sPAcon(j,lost).BusBusy(n,t,next,destfault,busy)
  )
+
  sum(j:NAT,
    rPAreq(j,immediate).
      ( BusBusy(n,t,invert(j,next),destfault,busy)
          <| not(get(j,next)) |> delta )
  )

SubactionGap(n:NAT,t:BoolTABLE,i:NAT)=
  BusIdle(n,t)
<| eq(i,n) |>
  sPDind(i,subactgap).SubactionGap(n,t,succ(i))

Resolve(n:NAT,t:BoolTABLE,next:BoolTABLE,i:NAT)=
(
  (
    ( sPAcon(i,won).sPCind(i).Resolve(n,t,next,succ(i)) )
  <| get(i,next) |>
    ( tau.Resolve(n,t,next,succ(i)) )
  )
<| lt(i,n) |>
  Resolve2(n,t,next)
)

Resolve2(n:NAT,t:BoolTABLE,next:BoolTABLE)=
(
  sum(j:NAT,
    rPDreq(j,End).
      (
        Resolve2(n,t,invert(j,next))
      <| get(j,next) |>
        delta
      )
  )
<| more(next) |>
  sum(j:NAT,
    sum(p:SIGNAL,
      rPDreq(j,p).
        (
          SubactionGap(n,t,0)
        <| is_end(p) |>
          Distribute(n,t,inita(n),inita(n),j,p,0)
        )
    )
  )
)

Distribute(n:NAT,
           t:BoolTABLE,
           next:BoolTABLE,
           destfault:BoolTABLE,
           busy:NAT,
           p:SIGNAL,
           i:NAT)=
(
  (
    (
      %% Signals can be handed over correctly
      ( sPDind(i,p).
          Distribute(n,t,next,destfault,busy,p,succ(i))
            <| or(not(is_header(p)),not(get(i,destfault))) |>
              delta )
    +
      %% Destination signals may be corrupted
      ( sum(dest:NAT,
          sPDind(i,sig(dest)).
            Distribute(n,t,next,invert(i,destfault),busy,p,succ(i))
        ) <| is_dest(p) |> delta )
    +
      %% Headers/Data/Acks may be corrupted
      ( sPDind(i,corrupt(p)).
          Distribute(n,t,next,destfault,busy,p,succ(i))
            <| is_hda(p) |> delta )
    +
      %% Headers/Data/Acks may get lost
      ( losesignal.Distribute(n,t,next,destfault,busy,p,succ(i))
          <| is_hda(p) |> delta )
    +
      %% Packets may be too large
      ( sPDind(i,p).sPDind(i,Dummy).
          Distribute(n,t,next,destfault,busy,p,succ(i))
            <| is_data(p) |> delta )
    +
      ( rPAreq(i,immediate).
          ( Distribute(n,t,invert(i,next),destfault,busy,p,i)
              <| not(get(i,next)) |> delta ) )
    )
  <| not(eq(i,busy)) |>
    tau.Distribute(n,t,next,destfault,busy,p,succ(i))
  )
<| lt(i,n) |>
  (
    BusBusy(n,t,next,destfault,n)
  <| is_end(p) |>
    BusBusy(n,t,next,destfault,busy)
  )
)

\end{lstlisting}
\SPACING


\subsection{The MAIN process in $\mu$CRL}

\begin{lstlisting}
act
  PDind,PDreq: NAT#SIGNAL
  PAcon: NAT#PAC
  PAreq: NAT#PAR
  PCind: NAT

comm
  rPDind|sPDind=PDind
  rPDreq|sPDreq=PDreq
  rPAcon|sPAcon=PAcon
  rPAreq|sPAreq=PAreq
  rPCind|sPCind=PCind

proc

 P1394(n:NAT)=
   hide({PDind, PDreq, PAcon, PAreq, PCind, arbresgap,losesignal},
     encap( {rPDind, sPDind, rPDreq, sPDreq, rPAcon,
             sPAcon, rPAreq, sPAreq, rPCind, sPCind},
         BUS(2) || LINK(2,0) || LINK(2,1)
       )
     )

% note: for 3 links, use BUS(3) || LINK(3,0) || LINK(3,1) || LINK(3,2), etc.

init P1394(2)

\end{lstlisting}
\SPACING


\section{Formal model in mCRL2}
\label{ANNEX-MCRL2}

\lstset{
  language=mCRL2,
  basicstyle=\ttfamily\small,
  columns=fullflexible, 
  commentstyle=\rmfamily\itshape,
  keywordstyle=\rmfamily\bfseries,
  xleftmargin=0pt
}


\subsection{Types and functions in mCRL2}

\begin{lstlisting}
sort CHECK = struct bottom | check;

sort DATA = struct d1 | d2;

map  crc : DATA -> CHECK;
eqn  crc(d1)=check;
     crc(d2)=check;

sort HEADER = struct h1 | h2;

map  crc : HEADER -> CHECK;
eqn  crc(h1)=check;
     crc(h2)=check;

sort ACK = struct a1 | a2;

map  crc : ACK -> CHECK;
eqn  crc(a1)=check;
     crc(a2)=check;

sort SIGNAL = struct sig(getdest:Nat) ? is_dest |
                     sig(gethead:HEADER,gethcrc:CHECK) ? is_header |
                     sig(getdata:DATA,getdcrc:CHECK) ? is_data |
                     sig(getack:ACK,getacrc:CHECK) ? is_ack |
                     Start ? is_start |
                     End ? is_end |
                     Prefix ? is_prefix |
                     subactgap ? is_sagap |
                     dhead ? is_dhead |
                     Dummy ? is_dummy;

map  is_physig,is_terminator : SIGNAL -> Bool;
     getcrc : SIGNAL -> CHECK;
var  s : SIGNAL;
eqn  is_physig(s) = is_start(s) || is_end(s) || is_prefix(s) || is_sagap(s);
     is_terminator(s)=is_end(s) || is_prefix(s);
     getcrc(s)=if(is_header(s),gethcrc(s),
               if(is_data(s),getdcrc(s),
               if(is_ack(s),getacrc(s),
                            bottom)));

map  is_hda : SIGNAL -> Bool;
     valid_hpart, valid_ack : SIGNAL -> Bool;
var  s : SIGNAL;
eqn  is_hda(s)=is_header(s) || is_data(s) || is_ack(s);
     valid_ack(s)=if(is_ack(s),getacrc(s)==check,false);
     valid_hpart(s)=if(is_header(s),gethcrc(s)==check,false);

map  corrupt : SIGNAL -> SIGNAL;
var  h : HEADER;
     d : DATA;
     a : ACK;
     c : CHECK;
eqn  corrupt(sig(h,c)) = sig(h,bottom);
     corrupt(sig(d,c)) = sig(d,bottom);
     corrupt(sig(a,c)) = sig(a,bottom);

sort SIG_TUPLE =
       struct quadruple (first:SIGNAL,
                         second:SIGNAL,
                         third:SIGNAL,
                         fourth:SIGNAL)
        | void ? is_void;

sort PAR = struct fair | immediate;

sort PAC = struct won | lost;

sort LDC = struct ackrec(ACK)
            |     ackmiss
            |     broadsent;

sort LDI = struct good (HEADER,DATA)
            |     broadrec (HEADER,DATA)
            |     dcrc_err (HEADER);

sort BOC = struct release | hold;

\end{lstlisting}
\SPACING


\subsection{The LINK process in mCRL2}

\begin{lstlisting}
act
  LDreq : Nat#Nat#HEADER#DATA;
  LDcon : Nat#LDC;
  LDind : Nat#LDI;
  LDres : Nat#ACK#BOC;

  sPDreq,rPDind : Nat#SIGNAL;
  sPAreq : Nat#PAR;
  rPAcon : Nat#PAC;
  rPCind : Nat;

proc LINK(n:Nat,i:Nat)=Link0(n,i,void);

     Link0(n:Nat,id:Nat,buffer:SIG_TUPLE)=
        is_void(buffer) ->
           ( sum dest:Nat,h:HEADER,d:DATA.
                ( dest<=n) -> LDreq(id,dest,h,d).
                     Link0(n,id,quadruple(dhead,
                                 sig(dest),
                                 sig(h,crc(h)),
                                 sig(d,crc(d))))<>delta) <>
              sPAreq(id,fair).Link1(n,id,buffer) +
        sum p:SIGNAL.
           rPDind(id,p).
              (is_start(p) -> Link4(n,id,buffer) <> Link0(n,id,buffer));

     Link1(n:Nat,id:Nat,p:SIG_TUPLE)=
        rPAcon(id,won).Link2req(n,id,p) +
        rPAcon(id,lost).Link0(n,id,p);

     Link2req(n:Nat,id:Nat,p:SIG_TUPLE)=
        rPCind(id).sPDreq(id,Start).
        rPCind(id).sPDreq(id,first(p)).
        rPCind(id).sPDreq(id,second(p)) .
        rPCind(id).sPDreq(id,third(p)).
        rPCind(id).sPDreq(id,fourth(p)).
        rPCind(id).sPDreq(id,End).
        ( (getdest(second(p))==n) ->
            LDcon(id,broadsent).Link0(n,id,void) <>
            Link3(n,id,void));

     Link3(n:Nat,id:Nat,buffer:SIG_TUPLE)=
        sum p:SIGNAL.
           rPDind(id,p).
           ( is_prefix(p) -> Link3(n,id,buffer) <>
           ( is_start(p) ->  Link3RA(n,id,buffer) <>
           ( is_sagap(p) ->  LDcon(id,ackmiss).Link0(n,id,buffer) <>
                             LDcon(id,ackmiss).LinkWSA(n,id,buffer,n)
           )));

     Link3RA(n:Nat,id:Nat,buffer:SIG_TUPLE)=
        sum a:SIGNAL.
           rPDind(id,a).
           ( is_sagap(a) ->  LDcon(id,ackmiss).Link0(n,id,buffer) <>
             ( is_physig(a) -> LDcon(id,ackmiss).LinkWSA(n,id,buffer,n) <>
                               Link3RE(n,id,buffer,a)));

     Link3RE(n:Nat,id:Nat,buffer:SIG_TUPLE,a:SIGNAL)=
        sum e:SIGNAL.
           rPDind(id,e).
           ((valid_ack(a) && is_terminator(e)) ->
                   LDcon(id,ackrec(getack(a))).LinkWSA(n,id,buffer,n) <>
             ( is_sagap(e) ->
                   LDcon(id,ackmiss).Link0(n,id,buffer) <>
                   LDcon(id,ackmiss).LinkWSA(n,id,buffer,n)
           ) );

     Link4(n:Nat,id:Nat,buffer:SIG_TUPLE)=
        sum dh:SIGNAL.
           rPDind(id,dh).
           ( is_physig(dh) ->
             ( is_sagap(dh) ->
                Link0(n,id,buffer) <>
                LinkWSA(n,id,buffer,n)) <>
             Link4DH(n,id,buffer));

     Link4DH(n:Nat,id:Nat,buffer:SIG_TUPLE)=
        sum dest:SIGNAL.rPDind(id,dest).
           ( is_dest(dest) ->
             ( (getdest(dest)==id) ->
                  sPAreq(id,immediate).Link4RH(n,id,buffer,id) <>
                  ( (getdest(dest)==n) ->
                      Link4RH(n,id,buffer,n) <>
                      LinkWSA(n,id,buffer,n)
                  )
             ) <>
             ( is_sagap(dest) ->
                  Link0(n,id,buffer) <>
                  LinkWSA(n,id,buffer,n)
           ) );

     Link4RH(n:Nat,id:Nat,buffer:SIG_TUPLE,dest:Nat)=
        sum h:SIGNAL.rPDind(id,h).
           ( valid_hpart(h) ->
                Link4RD(n,id,buffer,dest,h) <>
                LinkWSA(n,id,buffer,dest)
           );

     Link4RD(n:Nat,id:Nat,buffer:SIG_TUPLE,dest:Nat,h:SIGNAL)=
        sum d:SIGNAL.
           rPDind(id,d).
             ( is_data(d) ->
                  Link4RE(n,id,buffer,dest,h,d) <>
                  LinkWSA(n,id,buffer,dest)
             );

     Link4RE(n,id:Nat,buffer:SIG_TUPLE,dest:Nat,h:SIGNAL,d:SIGNAL)=
        sum e:SIGNAL.
           rPDind(id,e).
             ( is_terminator(e) ->
                  ( (dest==id) ->
                       Link4DRec(n,id,buffer,h,d) <>
                       Link4BRec(n,id,buffer,h,d)
                  ) <>
                  LinkWSA(n,id,buffer,dest)
             );

     Link4DRec(n:Nat,id:Nat,buffer:SIG_TUPLE,h:SIGNAL,d:SIGNAL)=
        (getcrc(d)==check) ->
           LDind(id,good(gethead(h),getdata(d))).rPAcon(id,won).Link5(n,id,buffer)
           <>
           LDind(id,dcrc_err(gethead(h))).rPAcon(id,won).Link5(n,id,buffer);

     Link4BRec(n:Nat,id:Nat,buffer:SIG_TUPLE,h:SIGNAL,d:SIGNAL)=
        (getcrc(d)==check) ->
           LDind(id,broadrec(gethead(h),getdata(d))).Link0(n,id,buffer) <>
           Link0(n,id,buffer);

     Link5(n,id:Nat,buffer:SIG_TUPLE)=
        sum a:ACK,b:BOC.LDres(id,a,b).Link6(n,id,buffer,sig(a,crc(a)),b) +
        rPCind(id).sPDreq(id,Prefix).Link5(n,id,buffer);

     Link6(n:Nat,id:Nat,buffer:SIG_TUPLE,p:SIGNAL,b:BOC)=
        rPCind(id).sPDreq(id,Start).rPCind(id).sPDreq(id,p).rPCind(id).
           ( (b==release) ->
                sPDreq(id,End).Link0(n,id,buffer) <>
                sPDreq(id,Prefix).Link7(n,id,buffer)
           );

     Link7(n,id:Nat,buffer:SIG_TUPLE)=
        rPCind(id).sPDreq(id,Prefix).Link7(n,id,buffer) +
        sum dest:Nat,h:HEADER,d:DATA. (dest<=n) ->
             LDreq(id,dest,h,d). Link2resp(n,id,buffer,
                      quadruple(dhead,sig(dest),sig(h,crc(h)),sig(d,crc(d))))<>delta;

     Link2resp(n:Nat,id:Nat,buffer:SIG_TUPLE,p:SIG_TUPLE)=
        rPCind(id).sPDreq(id,Start).
        rPCind(id).sPDreq(id,first(p)).
        rPCind(id).sPDreq(id,second(p)).
        rPCind(id).sPDreq(id,third(p)).
        rPCind(id).sPDreq(id,fourth(p)).
        rPCind(id).sPDreq(id,End).
        ( (getdest(second(p))==n) ->
             LDcon(id,broadsent).Link0(n,id,buffer) <>
             Link3(n,id,buffer)
        );

     LinkWSA(n:Nat,id:Nat,buffer:SIG_TUPLE,dest:Nat)=
        sum p:SIGNAL.rPDind(id,p).
           ( is_sagap(p) ->
                Link0(n,id,buffer) <>
                LinkWSA(n,id,buffer,dest)
           ) +
        (dest==id) -> rPAcon(id,won).rPCind(id).sPDreq(id,End).Link0(n,id,buffer)<>delta;

\end{lstlisting}
\SPACING


\subsection{The BUS process in mCRL2}

\begin{lstlisting}
sort BoolTABLE = List(struct pair(Nat,getbool:Bool));

map  inita : Nat -> BoolTABLE;
     invert : Nat#BoolTABLE -> BoolTABLE;
     get : Nat#BoolTABLE -> Bool;
var  n,m : Nat;
     b : Bool;
     bt1,bt2 : BoolTABLE;
eqn  inita(0)=[];
     n>0 -> inita(n)=pair(Int2Nat(n-1),false)|>inita(Int2Nat(n-1));

     invert(n,[])=[];
     invert(n,pair(m,b)|>bt1)=
        if(n==m,pair(m,!b)|>bt1,pair(m,b)|>invert(n,bt1));
     get(n,[])=false;
     get(n,pair(m,b)|>bt1)=if(n==m,b,get(n,bt1));

map  zero,one,more: BoolTABLE -> Bool;
var  n  : Nat;
     bt : BoolTABLE;
eqn  zero([])=true;
     zero(pair(n,true)|>bt)=false;
     zero(pair(n,false)|>bt)=zero(bt);
     one([])=false;
     one(pair(n,true)|>bt)=zero(bt);
     one(pair(n,false)|>bt)=one(bt);
     more(bt)=!zero(bt) && !one(bt);

act  rPAreq: Nat#PAR;
     rPDreq,sPDind: Nat#SIGNAL;
     sPAcon: Nat#PAC;
     sPCind: Nat;
     arbresgap;
     losesignal;
     internal;

proc BUS(n:Nat)=BusIdle(n, inita(n));

     BusIdle(n:Nat,t:BoolTABLE)=
        sum id:Nat,astat:PAR.(id<=n) ->
           rPAreq(id,astat).DecideIdle(n,t,id,astat)<>delta +
           !zero(t)->arbresgap.BusIdle(n,inita(n))<>delta;

     DecideIdle(n:Nat,t:BoolTABLE,id:Nat,astat:PAR)=
        (!get(id,t)) ->
          sPAcon(id,won).BusBusy(n,invert(id,t),inita(n),inita(n),id) <>
          sPAcon(id,lost).BusIdle(n,t);

     BusBusy(n:Nat,t,next,destfault:BoolTABLE,busy:Nat)=
        (busy<n) ->
           ( sPCind(busy).
                (sum p:SIGNAL.rPDreq(busy,p).Distribute(n,t,next,destfault,busy,p,0))
           ) <>
           ( zero(next) ->
                SubactionGap(n,t,0) <>
                  Resolve(n,t,next,0)
           ) +
       sum j:Nat.(j<=n) ->
          rPAreq(j,fair).sPAcon(j,lost).BusBusy(n,t,next,destfault,busy)<>delta +
       sum j:Nat.(j<=n) -> rPAreq(j,immediate).
          (!get(j,next) -> BusBusy(n,t,invert(j,next),destfault,busy)<>delta)<>delta;

     SubactionGap(n:Nat,t:BoolTABLE,i:Nat)=
        (i==n) ->
           BusIdle(n,t) <>
           sPDind(i,subactgap).SubactionGap(n,t,i+1);

     Resolve(n:Nat,t,next:BoolTABLE,i:Nat)=
        (i<n) ->
        (get(i,next) ->
            sPAcon(i,won).sPCind(i).Resolve(n,t,next,i+1) <>
            internal.Resolve(n,t,next,i+1)
        ) <>
        Resolve2(n,t,next);

     Resolve2(n:Nat,t:BoolTABLE,next:BoolTABLE)=
        more(next) ->
           (sum j:Nat.(j<=n) -> rPDreq(j,End).(get(j,next) ->
           Resolve2(n,t,invert(j,next))<>delta)<>delta) <>
           (sum j:Nat,p:SIGNAL.(j<=n) ->
              rPDreq(j,p).
                 (is_end(p) ->
                     SubactionGap(n,t,0) <>
                     Distribute(n,t,inita(n),inita(n),j,p,0)
           )<>delta);

     Distribute(n:Nat,t,next,destfault:BoolTABLE,busy:Nat,p:SIGNAL,i:Nat)=
        (i<n) ->
        ( (i!=busy) ->
          ( %% Signals can be handed over correctly
            (!is_header(p) || !get(i,destfault)) ->
               sPDind(i,p).Distribute(n,t,next,destfault,busy,p,i+1)<>delta +
            %% Destination signals may be corrupted
            sum dest:Nat.(is_dest(p) && dest<=n) ->
               sPDind(i,sig(dest)).
                  Distribute(n,t,next,invert(i,destfault),busy,p,i+1)<>delta +
            %% Headers/Data/Acks may be corrupted
            is_hda(p) ->
               sPDind(i,corrupt(p)).
                  Distribute(n,t,next,destfault,busy,p,i+1)<>delta +
            %% Headers/Data/Acks may get lost
            is_hda(p) ->
               losesignal.Distribute(n,t,next,destfault,busy,p,i+1)<>delta +
            %% Packets may be too large
            is_data(p) ->
               sPDind(i,p).sPDind(i,Dummy).
                  Distribute(n,t,next,destfault,busy,p,i+1)<>delta +
            (!get(i,next)) ->
               rPAreq(i,immediate).
                  Distribute(n,t,invert(i,next),destfault,busy,p,i)<>delta
         ) <>
         %% i==busy
         internal.Distribute(n,t,next,destfault,busy,p,i+1)
       ) <>
       %% i>=n
       ( is_end(p) ->
            BusBusy(n,t,next,destfault,n) <>
            BusBusy(n,t,next,destfault,busy)
       );

\end{lstlisting}
\SPACING


\subsection{The MAIN process in mCRL2}

\begin{lstlisting}
act
  cPDreq,cPDind : Nat#SIGNAL;
  cPAreq : Nat#PAR;
  cPAcon : Nat#PAC;
  cPCind : Nat;

proc P1394(n:Nat)=
        allow({LDreq,LDcon,LDind,LDres},
           hide({arbresgap,losesignal,internal,cPDind,cPDreq,cPAcon,cPAreq,cPCind},
              comm({rPDind|sPDind->cPDind,rPDreq|sPDreq->cPDreq,rPAcon|sPAcon->cPAcon,
                    rPAreq|sPAreq->cPAreq,rPCind|sPCind->cPCind},
                  allow({LDreq,LDcon,LDind,LDres,arbresgap,losesignal,internal,
                              rPDind|sPDind,rPDreq|sPDreq,rPAcon|sPAcon,
                    rPAreq|sPAreq,rPCind|sPCind},
                          BUS(2) || LINK(2,0) || LINK(2,1)))));

% note: for 3 links, use BUS(3) || LINK(3,0) || LINK(3,1) || LINK(3,2), etc.

init P1394(2);

\end{lstlisting}
\SPACING


\section{Formal model in LOTOS}
\label{ANNEX-LOTOS}

\lstset{
  language=LOTOS,
  basicstyle=\ttfamily\small,
  columns=fullflexible, 
  commentstyle=\rmfamily\itshape,
  keywordstyle=\rmfamily\bfseries,
  xleftmargin=0pt
}


\subsection{Types and functions in LOTOS}

\begin{lstlisting}
type CHECK is Boolean
   sorts
      CHECK
   opns
      bottom (*! constructor *) : -> CHECK
      check (*! constructor *) : -> CHECK
      eq : CHECK, CHECK -> Bool
   eqns
      forall x, y : CHECK
      ofsort Bool
         eq (x, x) = true;
         (* otherwise *) eq (x, y) = false;
endtype

(*---------------------------------------------------------------------------*)

type DATA is CHECK
   sorts
      DATA
   opns
      d1 (*! constructor *) : -> DATA
      (* for verification, this type is restricted to a single value *)
      (* d2 {*! constructor *} : -> DATA *)
      crc : DATA -> CHECK
      eq : DATA, DATA -> Bool
   eqns
      forall x, y : DATA
      ofsort Bool
         eq (x, x) = true;
         (* otherwise *) eq (x, y) = false;
      ofsort CHECK
         crc (x) = check;
endtype

(*---------------------------------------------------------------------------*)

type HEADER is CHECK
   sorts
      HEADER
   opns
      h1 (*! constructor *) : -> HEADER
      (* for verification, this type is restricted to a single value *)
      (* h2 {*! constructor *} : -> HEADER *)
      crc : HEADER -> CHECK
      eq : HEADER, HEADER -> Bool
   eqns
      forall x, y : HEADER
      ofsort Bool
         eq (x, x) = true;
         (* otherwise *) eq (x, y) = false;
      ofsort CHECK
         crc (x) = check;
endtype

(*---------------------------------------------------------------------------*)

type ACK is CHECK
   sorts
      ACK
   opns
      a1 (*! constructor *) : -> ACK
      (* for verification, this type is restricted to a single value *)
      (* a2 {*! constructor *} : -> ACK *)
      crc : ACK -> CHECK
      eq : ACK, ACK -> Bool
   eqns
      forall x, y : ACK
      ofsort Bool
         eq (x, x) = true;
         (* otherwise *) eq (x, y) = false;
      ofsort CHECK
         crc (x) = check;
endtype

(*---------------------------------------------------------------------------*)

type BOC is CHECK
   sorts
      BOC
   opns
      release (*! constructor *),
      hold (*! constructor *),
      no_op (*! constructor *) : -> BOC
      eq : BOC, BOC -> Bool
   eqns
      forall x, y : BOC
      ofsort Bool
         eq (x, x) = true;
         (* otherwise *) eq (x, y) = false;
endtype

(*---------------------------------------------------------------------------*)

type PHY_AREQ is CHECK
   sorts
      PHY_AREQ
   opns
      fair (*! constructor *),
      immediate (*! constructor *) : -> PHY_AREQ
      eq : PHY_AREQ, PHY_AREQ -> Bool
   eqns
     forall x, y : PHY_AREQ
     ofsort Bool
         eq (x, x) = true;
         (* otherwise *) eq (x, y) = false;
endtype

(*---------------------------------------------------------------------------*)

type PHY_ACONF is CHECK
   sorts
      PHY_ACONF
   opns
      won (*! constructor *),
      lost (*! constructor *) : -> PHY_ACONF
      eq : PHY_ACONF, PHY_ACONF -> Bool
   eqns
     forall x, y : PHY_ACONF
     ofsort Bool
         eq (x, x) = true;
         (* otherwise *) eq (x, y) = false;
endtype

(*---------------------------------------------------------------------------*)

type SIGNAL is ACK, CHECK, DATA, HEADER, NaturalNumber
   sorts
      SIGNAL
   opns
      destsig (*! constructor *) : Nat -> SIGNAL
      headsig (*! constructor *) : HEADER, CHECK -> SIGNAL
      datasig (*! constructor *) : DATA, CHECK -> SIGNAL
      acksig (*! constructor *) : ACK, CHECK -> SIGNAL
      dhead (*! constructor *) : -> SIGNAL
      Start (*! constructor *) : -> SIGNAL
      End (*! constructor *) : -> SIGNAL
      Prefix (*! constructor *) : -> SIGNAL
      subactgap (*! constructor *) : -> SIGNAL
      Dummy (*! constructor *) : -> SIGNAL
      is_dest, is_header, is_data, is_ack, is_physig : SIGNAL -> Bool
      valid_hpart, valid_ack : SIGNAL -> Bool
      getdest : SIGNAL -> Nat
      getdcrc : SIGNAL -> CHECK
      getdata : SIGNAL -> DATA
      gethead : SIGNAL -> HEADER
      getack : SIGNAL -> ACK
      corrupt : SIGNAL -> SIGNAL
      eq : SIGNAL, SIGNAL -> Bool
   eqns
      forall n : Nat, c : CHECK, h : HEADER, d : DATA, a : ACK, s, s1, s2 : SIGNAL
      ofsort Bool
         is_dest (destsig (n)) = true;
         (* otherwise *) is_dest (s) = false;
         is_header (headsig (h, c)) = true;
         (* otherwise *) is_header (s) = false;
         is_data (datasig (d, c)) = true;
         (* otherwise *) is_data (s) = false;
         is_ack (acksig (a, c)) = true;
         (* otherwise *) is_ack (s) = false;
         is_physig (Start) = true;
         is_physig (End) = true;
         is_physig (Prefix) = true;
         is_physig (subactgap) = true;
         (* otherwise *) is_physig (s) = false;
         valid_ack (acksig (a, c)) = eq (c, check);
         (* otherwise *) valid_ack (s) = false;
         valid_hpart (headsig (h, c)) = eq (c, check);
         (* otherwise *) valid_hpart (s) = false;
      ofsort Nat
         getdest (destsig (n)) = n;
         (* otherwise getdest (s) is undefined *)
      ofsort HEADER
         gethead (headsig (h, c)) = h;
         (* otherwise gethead (s) is undefined *)
      ofsort CHECK
         getdcrc (datasig (d, c)) = c;
         (* otherwise getdcrc (s) is undefined *)
      ofsort DATA
         getdata (datasig (d, c)) = d;
         (* otherwise getdata (s) is undefined *)
      ofsort ACK
         getack (acksig (a, c)) = a;
         (* otherwise getack (s) is undefined *)
      ofsort SIGNAL
         corrupt (headsig (h, c)) = headsig (h, bottom);
         corrupt (datasig (d, c)) = datasig (d, bottom);
         corrupt (acksig (a, c)) = acksig (a, bottom);
      ofsort Bool
         eq (s1, s1) = true;
         (* otherwise *) eq (s1, s2) = false;
endtype

(*---------------------------------------------------------------------------*)

type SIG_TUPLE is Boolean, SIGNAL
   sorts
      SIG_TUPLE
   opns
      quadruple (*! constructor *) : SIGNAL, SIGNAL, SIGNAL, SIGNAL -> SIG_TUPLE
      void (*! constructor *) : -> SIG_TUPLE
      first, second, third, fourth : SIG_TUPLE -> SIGNAL
      is_void : SIG_TUPLE -> Bool
   eqns
      forall s1, s2, s3, s4 : SIGNAL
      ofsort SIGNAL
         first (quadruple (s1, s2, s3, s4)) = s1;
         second (quadruple (s1, s2, s3, s4)) = s2;
         third (quadruple (s1, s2, s3, s4)) = s3;
         fourth (quadruple (s1, s2, s3, s4)) = s4;
      ofsort Bool
         is_void (void) = true;
         is_void (quadruple (s1, s2, s3, s4)) = false;
endtype

(*---------------------------------------------------------------------------*)

type LIN_DCONF is ACK
   sorts
      LIN_DCONF
   opns
      ackrec (*! constructor *) : ACK -> LIN_DCONF
      ackmiss (*! constructor *),
      broadsent (*! constructor *) : -> LIN_DCONF
endtype

(*---------------------------------------------------------------------------*)

type LIN_DIND is Boolean, DATA, HEADER
   sorts
      LIN_DIND
   opns
      good (*! constructor *),
      broadrec (*! constructor *) : HEADER, DATA -> LIN_DIND
      dcrc_err (*! constructor *) : HEADER -> LIN_DIND
      is_broadrec : LIN_DIND -> Bool
   eqns
      forall h: HEADER, d: DATA, xind: LIN_DIND
      ofsort Bool
         is_broadrec (broadrec (h, d)) = true;
         (* otherwise *) is_broadrec (xind) = false;
endtype

(*---------------------------------------------------------------------------*)

type BoolTABLE is Boolean, NaturalNumber
   sorts
      BoolTABLE
   opns
      empty (*! constructor *) : -> BoolTABLE
      btable (*! constructor *) : Nat, Bool, BoolTABLE -> BoolTABLE
      init : Nat -> BoolTABLE
      invert : Nat, BoolTABLE -> BoolTABLE
      get : Nat, BoolTABLE -> Bool
      zero, one, more : BoolTABLE -> Bool
   eqns
      forall n, n1, n2 : Nat, b : Bool, t : BoolTABLE
      ofsort BoolTABLE
         init (0) = empty;
         init (Succ (n)) = btable (n, false, init (n));
         invert (n, empty) = empty;
         n1 eq n2 => invert (n1, btable (n2, b, t)) = btable (n2, not (b), t);
         n1 ne n2 => invert (n1, btable (n2, b, t)) = btable (n2, b, invert (n1, t));
      ofsort Bool
         (* get (n, empty) is undefined *)
         n1 eq n2 => get (n1, btable (n2, b, t)) = b;
         n1 ne n2 => get (n1, btable (n2, b, t)) = get (n1, t);
      ofsort Bool
         zero (empty) = true;
         zero (btable (n, true, t)) = false;
         zero (btable (n, false, t)) = zero (t);
         one (empty) = false;
         one (btable (n, true, t)) = zero (t);
         one (btable (n, false, t)) = one (t);
         more (t) = not (zero (t)) and not (one (t));
endtype

(*---------------------------------------------------------------------------*)

type Version is
   sorts
      Version
   opns
      ko (*! constructor *),
      ok (*! constructor *) : -> Version
endtype

(*---------------------------------------------------------------------------*)

type Scenario is Boolean, Natural
   sorts
      Scenario
   opns
      scenario_1 (*! constructor *),
      scenario_2 (*! constructor *),
      scenario_3_2 (*! constructor *),
      scenario_3_3 (*! constructor *),
      scenario_3_4 (*! constructor *) : -> Scenario
      _eq_ : Scenario, Scenario -> Bool
   eqns
      forall s1, s2: Scenario
      ofsort Bool
         s1 eq s1 = true;
         (* otherwise *) s1 eq s2 = false;
endtype

\end{lstlisting}
\SPACING


\subsection{The LINK process in LOTOS}

\lstinputlisting{spec/LOTOS/LINK.lib}
\SPACING


\subsection{The BUS process in LOTOS}

\begin{lstlisting}

process Bus [PAreq, PDreq, PDind, PAcon, PCind, arbresgap, losesignal]
             (n: Nat) : noexit :=
   BusIdle [PAreq, PDreq, PDind, PAcon, PCind, arbresgap, losesignal] (n, init (n))
endproc

(* ------------------------------------------------------------------------- *)

process BusIdle [PAreq, PDreq, PDind, PAcon, PCind, arbresgap, losesignal]
                 (n: Nat, t: BoolTABLE) : noexit :=
   PAreq ?id: Nat ?astat: PHY_AREQ [id lt n];
   DecideIdle [PAreq, PDreq, PDind, PAcon, PCind, arbresgap, losesignal] (n, t, id,
                astat)
   []
   [not (zero(t))] ->
      arbresgap;
      BusIdle [PAreq, PDreq, PDind, PAcon, PCind, arbresgap, losesignal] (n, init (n))
endproc

(* ------------------------------------------------------------------------- *)

process DecideIdle [PAreq, PDreq, PDind, PAcon, PCind, arbresgap, losesignal]
                    (n: Nat, t: BoolTABLE, id: Nat, astat: PHY_AREQ) : noexit :=
   [get (id, t) eq false] ->
      PAcon !id !won;
      BusBusy [PAreq, PDreq, PDind, PAcon, PCind, arbresgap, losesignal] (n,
                invert (id, t), init (n), init (n), id)
   []
   [get (id, t) eq true] ->
      PAcon !id !lost;
      BusIdle [PAreq, PDreq, PDind, PAcon, PCind, arbresgap, losesignal] (n, t)
endproc

(* ------------------------------------------------------------------------- *)

process BusBusy [PAreq, PDreq, PDind, PAcon, PCind, arbresgap, losesignal]
                 (n: Nat, t: BoolTABLE, next: BoolTABLE, destfault: BoolTABLE,
                 busy: Nat) : noexit :=
   [busy lt n] ->
      PCind !busy;
      PDreq !busy ?p: SIGNAL;
      Distribute [PAreq, PDreq, PDind, PAcon, PCind, arbresgap, losesignal]
                  (n, t, next, destfault, busy, p, 0)
   []
   [not (busy lt n)] ->
       (
       [zero (next)] ->
          SubactionGap [PAreq, PDreq, PDind, PAcon, PCind, arbresgap, losesignal]
                         (n, t, 0)
       []
       [not (zero (next))] ->
           Resolve [PAreq, PDreq, PDind, PAcon, PCind, arbresgap, losesignal]
                    (n, t, next, 0)
       )
   []
   PAreq ?j: Nat !fair [j lt n];
   PAcon !j !lost;
   BusBusy [PAreq, PDreq, PDind, PAcon, PCind, arbresgap, losesignal]
            (n, t, next, destfault, busy)
   []
   PAreq ?j: Nat !immediate [not (get (j, next)) and (j lt n)];
   BusBusy [PAreq, PDreq, PDind, PAcon, PCind, arbresgap, losesignal]
            (n, t, invert (j, next), destfault, busy)
endproc

(* ------------------------------------------------------------------------- *)

process SubactionGap [PAreq, PDreq, PDind, PAcon, PCind, arbresgap, losesignal]
                      (n: Nat, t: BoolTABLE, j: Nat) : noexit :=
   [j eq n] ->
      BusIdle [PAreq, PDreq, PDind, PAcon, PCind, arbresgap, losesignal] (n, t)
   []
   [j ne n] ->
      PDind !j !subactgap;
      SubactionGap [PAreq, PDreq, PDind, PAcon, PCind, arbresgap, losesignal]
                    (n, t, succ (j))
endproc

(* ------------------------------------------------------------------------- *)

process Resolve [PAreq, PDreq, PDind, PAcon, PCind, arbresgap, losesignal]
                 (n: Nat, t: BoolTABLE, next: BoolTABLE, j: Nat) : noexit :=
   [j lt n] ->
      (
      [get (j, next) eq true] ->
         PAcon !j !won;
         PCind !j;
         Resolve [PAreq, PDreq, PDind, PAcon, PCind, arbresgap, losesignal]
                  (n, t, next, succ (j))
      []
      [get (j, next) eq false] ->
         Resolve [PAreq, PDreq, PDind, PAcon, PCind, arbresgap, losesignal]
                  (n, t, next, succ (j))
      )
   []
   [not (j lt n)] ->
      Resolve2 [PAreq, PDreq, PDind, PAcon, PCind, arbresgap, losesignal]
                (n, t, next)
endproc

(* ------------------------------------------------------------------------- *)

process Resolve2 [PAreq, PDreq, PDind, PAcon, PCind, arbresgap, losesignal]
                  (n: Nat, t: BoolTABLE, next: BoolTABLE) : noexit :=
   [more (next)] ->
      PDreq ?j: Nat !End [get (j, next) and (j lt n)];
      Resolve2 [PAreq, PDreq, PDind, PAcon, PCind, arbresgap, losesignal]
                (n, t, invert (j, next))
   []
   [not (more (next))] ->
      PDreq ?j: Nat ?p: SIGNAL [j lt n];
      (
      [eq (p, End)] ->
         SubactionGap [PAreq, PDreq, PDind, PAcon, PCind, arbresgap, losesignal]
                        (n, t, 0)
      []
      [not (eq (p, End))] ->
         Distribute [PAreq, PDreq, PDind, PAcon, PCind, arbresgap, losesignal]
                     (n, t, init (n), init (n), j, p, 0)
      )
endproc

(* ------------------------------------------------------------------------- *)

process Distribute [PAreq, PDreq, PDind, PAcon, PCind, arbresgap, losesignal]
                    (n: Nat, t, next, destfault: BoolTABLE, busy: Nat, p: SIGNAL,
                    j: Nat) : noexit :=
   [j lt n] ->
      (
      [j ne busy] ->
         (
         [not (is_header (p)) or not (get (j, destfault))] ->
            PDind !j !p;
            Distribute [PAreq, PDreq, PDind, PAcon, PCind, arbresgap, losesignal]
                        (n, t, next, destfault, busy, p, succ (j))
         []
         [is_dest (p)] ->
            (
            choice dest: Nat []
               PDind !j !destsig (dest);
               Distribute [PAreq, PDreq, PDind, PAcon, PCind, arbresgap, losesignal]
                           (n, t, next, invert (j, destfault), busy, p, succ (j))
            )
         []
         [is_header (p) or (is_data (p) or is_ack (p))] ->
            PDind !j !corrupt (p);
            Distribute [PAreq, PDreq, PDind, PAcon, PCind, arbresgap, losesignal]
                        (n, t, next, destfault, busy, p, succ (j))
         []
         [is_header (p) or (is_data (p) or is_ack (p))] ->
            losesignal;
            Distribute [PAreq, PDreq, PDind, PAcon, PCind, arbresgap, losesignal]
                        (n, t, next, destfault, busy, p, succ (j))
         []
         [is_data (p)] ->
            PDind !j !p;
            PDind !j !Dummy;
            Distribute [PAreq, PDreq, PDind, PAcon, PCind, arbresgap, losesignal]
                        (n, t, next, destfault, busy, p, succ (j))
         []
         PAreq !j !immediate [not (get (j, next))];
            Distribute [PAreq, PDreq, PDind, PAcon, PCind, arbresgap, losesignal]
                        (n, t, invert (j, next), destfault, busy, p, j)
         )
      []
      [j eq busy] ->
         Distribute [PAreq, PDreq, PDind, PAcon, PCind, arbresgap, losesignal]
                     (n, t, next, destfault, busy, p, succ (j))
      )
   []
   [not (j lt n)] ->
      (
      [eq (p, End)] ->
         BusBusy [PAreq, PDreq, PDind, PAcon, PCind, arbresgap, losesignal]
                  (n, t, next, destfault, n)
      []
      [not (eq (p, End))] ->
         BusBusy [PAreq, PDreq, PDind, PAcon, PCind, arbresgap, losesignal]
                  (n, t, next, destfault, busy)
      )
endproc

\end{lstlisting}
\SPACING


\subsection{The TRANS process in LOTOS}

\begin{lstlisting}
process Trans [LDreq, LDcon, LDind, LDres, TDreq] (n, id: Nat, v: Version) : noexit :=
   hide TX0 in
      (
      TransReq [LDreq, LDcon, TDreq, TX0] (n, id)
      |[TX0]|
      TransRes [LDind, LDres, TX0] (id, v)
      )
endproc

(*---------------------------------------------------------------------------*)

process TransReq [LDreq, LDcon, TDreq, TX0] (n, id: Nat) : noexit :=
   TDreq !id ?dest: Nat ?h: HEADER ?d: DATA [dest le n];
   (
   TX0;
   exit (dest, h, d)
   []
   exit (dest, h, d)
   ) >> accept dest: Nat, h: HEADER, d: DATA in
   (
   LDreq !id !dest !h !d;
      (
      [dest eq n] ->
         LDcon !id !broadsent;
         TransReq [LDreq, LDcon, TDreq, TX0] (n, id)
      []
      [dest ne n] ->
         (
         choice a: ACK []
            LDcon !id !ackrec (a);
            TransReq [LDreq, LDcon, TDreq, TX0] (n, id)
         )
      []
         LDcon !id !ackmiss;
         TransReq [LDreq, LDcon, TDreq, TX0] (n, id)
      )
   )
endproc

(*---------------------------------------------------------------------------*)

process TransRes [LDind, LDres, TX0] (id: Nat, v: Version) : noexit :=
   LDind !id ?l: LIN_DIND;
   (
   [is_broadrec (l)] ->
      (
      [v = ko] ->
         (* original (incorrect) specification *)
         LDres !id !a1 !no_op;
         TransRes [LDind, LDres, TX0] (id, v)
      []
      [v = ok] ->
         (* correct specification *)
         TransRes [LDind, LDres, TX0] (id, v)
      )
   []
   [not (is_broadrec (l))] ->
      (
      choice a: ACK []
         (
         (* concatenated response = lock transaction *)
         TX0;
         LDres !id !a !hold;
         TransRes [LDind, LDres, TX0] (id, v)
         []
         (* split response *)
         LDres !id !a !release;
         TransRes [LDind, LDres, TX0] (id, v)
         )
      )
   )
endproc

\end{lstlisting}
\SPACING


\subsection{The APPLI process in LOTOS}

\lstinputlisting{spec/LOTOS/APPLI.lib}
\SPACING


\subsection{The NODE process in LOTOS}

\begin{lstlisting}
process Node [LDreq, LDcon, LDind, LDres, PDreq, PDind, PAreq, PAcon, PCind]
             (n: Nat, id: Nat, v: Version, s: Scenario) : noexit :=
   hide TDreq in
      (
      Link [LDreq, LDcon, LDind, LDres, PDreq, PDind, PAreq, PAcon, PCind] (n, id)
      |[LDreq, LDcon, LDind, LDres]|
      Trans [LDreq, LDcon, LDind, LDres, TDreq] (n, id, v)
      |[TDreq]|
      Application [TDreq] (n, id, s)
      )
endproc

\end{lstlisting}
\SPACING


\subsection{The MAIN process in LOTOS}

\lstinputlisting{spec/LOTOS/scen3_orig_2_4.lotos}

For model-checking purposes, a complementary file restricts the set of
natural numbers, e.g., to the finite range $\{0, ..., 2\}$ in the above example.
\SPACING


\section{Formal model in LNT}
\label{ANNEX-LNT}

\lstset{
  language=LNT,
  basicstyle=\ttfamily\small,
  columns=fullflexible, 
  commentstyle=\rmfamily\itshape,
  keywordstyle=\rmfamily\bfseries,
  xleftmargin=0pt
}


\subsection{Types and functions in LNT}

\lstinputlisting{spec/LNT/DATA.lnt}
\SPACING


\subsection{Channels in LNT}

\begin{lstlisting}
module CHANNELS (DATA) is

channel Id is
   (n: Nat)
end channel

channel Sig is
   (id: Nat, flag: SIGNAL)
end channel

channel Areq is
   (id: Nat, flag: PHY_AREQ)
end channel

channel Acon is
   (id: Nat, flag: PHY_ACONF)
end channel

channel Ack is
   (id: Nat, a: ACK, b: BOC)
end channel

channel Dreq is
   (id: Nat, dest: Nat, h: HEADER, d: DATA)
end channel

channel Dind is
   (id: Nat, l: LIN_DIND)
end channel

channel Dcon is
   (id: Nat, l: LIN_DCONF)
end channel

end module

\end{lstlisting}
\SPACING


\subsection{The LINK process in LNT}

\lstinputlisting{spec/LNT/LINK.lnt}
\SPACING


\subsection{The BUS process in LNT}

\lstinputlisting{spec/LNT/BUS.lnt}
\SPACING


\subsection{The TRANS process in LNT}

\lstinputlisting{spec/LNT/TRANS.lnt}
\SPACING


\subsection{The APPLI process in LNT}

\lstinputlisting{spec/LNT/APPLI.lnt}
\SPACING


\subsection{The NODE process in LNT}

\lstinputlisting{spec/LNT/NODE.lnt}
\SPACING


\subsection{The MAIN process in LNT}

\lstinputlisting{spec/LNT/scen3_orig_2_4.lnt}

\end{document}